\documentclass[12pt,letterpaper]{article}
\usepackage[latin9]{inputenc}
\usepackage{amsmath}
\usepackage{amssymb}
\usepackage{graphicx}
\usepackage{setspace}
\usepackage{esint}
\usepackage[authoryear]{natbib}

\makeatletter

\DeclareFontEncoding{LGR}{}{}
\DeclareTextSymbol{\~}{LGR}{126}

\@ifundefined{date}{}{\date{}}
\newcommand{\qed}{\hfill$\blacksquare$\par}
\newcommand{\plim}{\stackrel{p}{\to}}

\newtheorem{define}{D{\small EFINITION}}
\newtheorem{lemma}{L{\small EMMA}}
\newtheorem{theorem}{T{\small HEOREM}}
\newtheorem{assumption}{A{\small SSUMPTION}}

\usepackage{amsfonts}
\usepackage{latexsym}
\usepackage{bbm}
\usepackage{subcaption}
\usepackage{varwidth}
\usepackage{caption}
\usepackage{lscape}
\usepackage{tablefootnote}
\usepackage{afterpage}

\makeatother

\begin{document}
\begin{singlespace}

\title{Simultaneous Selection of Optimal Bandwidths for the Sharp Regression Discontinuity Estimator\thanks{Earlier versions of this paper were titled ``Optimal Bandwidth Selection for Differences of Nonparametric Estimators with an Application to the Sharp Regression Discontinuity Design\char`\"{} and presented at Academia Sinica, the Japanese Economic Association Spring Meeting, LSE, the North American Winter Meeting of the Econometric Society,
UC Berkeley, UCL, and Yale. Valuable comments were received from seminar participants. We are especially grateful to Yoshihiko Nishiyama, Jack
Porter and Jim Powell for many helpful comments. We also thank Jens Ludwig and Douglas Miller for making the data used in \citet{lm07}
publicly available. Yoko Sakai provided expert research assistance. This research was supported by Grants-in-Aid for Scientific Research
No.\ 22243020 and No.\ 23330070 from the Japan Society for the Promotion of Science.}}
\end{singlespace}

\author{Yoichi Arai\thanks{National Graduate Institute for Policy Studies (GRIPS), 7-22-1 Roppongi,
Minato-ku, Tokyo 106-8677, Japan; yarai@grips.ac.jp} and Hidehiko Ichimura\thanks{Department of Economics, University of Tokyo, 7-3-1 Hongo, Bunkyo-ku,
Tokyo, 113-0033, Japan; ichimura@e.u-tokyo.ac.jp}}
\maketitle
\begin{abstract}
 A new bandwidth selection rule that
 uses different bandwidths for the local linear regression estimators on
 the left and the right of the cut-off point is proposed for the sharp
 regression discontinuity estimator of the mean program impact at the
 cut-off point.  The asymptotic mean squared error of the estimator using
 the proposed bandwidth selection rule is shown to be smaller than other
 bandwidth selection rules proposed in the literature.  An extensive
 simulation study shows that the proposed method's performances for the
 samples sizes 500, 2000, and 5000 closely match the theoretical predictions.
\end{abstract}
\begin{singlespace}
 {\em Key words}: Bandwidth selection, local linear regression,
regression discontinuity design
\end{singlespace}


\newpage{}

\section{Introduction}

The regression discontinuity (RD) is a quasi-experimental design to evaluate causal effects introduced by \citet{tc60} and developed by \citet{htv01}. 
A large number of empirical studies are carried out using the RD design in various areas of economics. See \citet{il08}, \citet{vdk08}, \citet{ll10} and \citet{dl11} for an overview and lists of empirical researches.\footnote{The RD approach has been extended in various directions. 
For example, \citet{clpw12} and \citet{dl14} examine how the RD estimate change when the discontinuity point change in the neighborhood of the RD
point and \citet{ffm12} considered the quantile treatment effect in the context of the RD design.}

We consider the sharp RD design in which whether
a value of the assignment variable exceeds a known cut-off value or not
determines the treatment status. 
A parameter of interest is the average treatment effect at the cut-off point. 
The average treatment effect is given by the difference between the two conditional mean functions at the cut-off point. 
This implies that estimating the treatment effect amounts to estimating two functions at the boundary point. One of the most frequently used estimation
methods is the local linear regression (LLR) because of its superior performance at the boundary. 
See \citet{fa92,fa93} and \citet{por03}.

A particular nonparametric estimator is undefined unless 
the smoothing parameter selection rule is specified, and it is well recognized that choosing an
appropriate smoothing parameter is a key implementation issue. 
In the RD setting, currently the most widely used method is developed by \citet{ik12} (hereafter IK). Other methods are the cross-validation
(\citet{lm05,lm07} (hereafter LM)) and the plug-in method (\citet{dm08}).
While the latter two approaches make use of the bandwidth selection rules that are tailored to estimating the regression functions, IK pays
attention to the essential fact that the parameter of interest is the difference of two conditional mean functions.\footnote{\citet{cct14} proposes robust confidence intervals for both sharp and fuzzy RD designs and \citet{mfl14} provide inferential procedures for the fuzzy RD design when identification is weak.}

\begin{figure}[h]
\vspace{5mm}
 \centering \includegraphics[width=12cm]{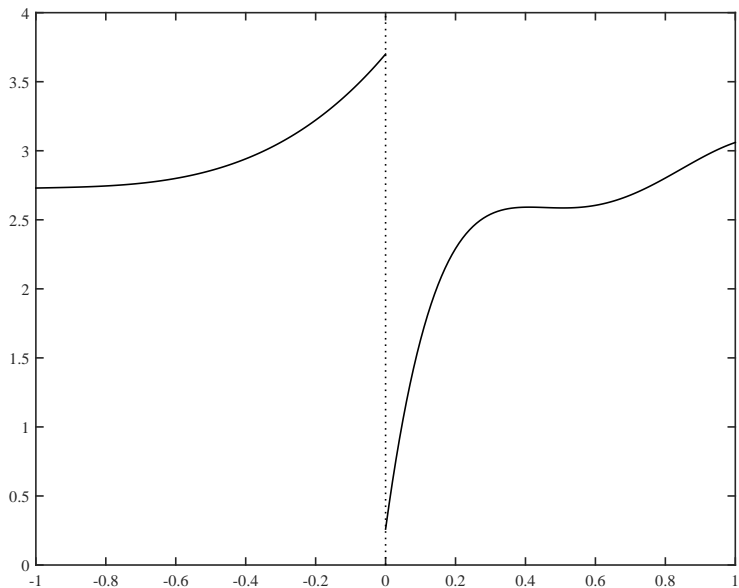} \protect\caption{Conditional mean functions of outcomes based on \citet{lm07}. The
line on the left of the cut-off point, zero, depicts the conditional
mean function of the potential outcome for untreated conditional on
the assignment variable. Similarly, the line on the right of the cut-off
point draws the corresponding function for treated.}
\label{fig:intro} 
\end{figure}

In the context of RD design, using two bandwidths for estimating two functions is a natural approach. 
The curvatures of the conditional mean functions for treated and untreated in the vicinity of the cut-off point may differ significantly. Figure \ref{fig:intro}
illustrates the situation motivated by \citet{lm07} where the cut-off value is depicted by a dotted vertical line. 
The solid lines depict two conditional mean functions to estimate. 
If we were to use a single bandwidth which is relatively large, it will
incur a large bias to estimate the conditional mean function on the
right of the cut-off point. 
On the other hand, using a single bandwidth which is relatively small
will lead to a smaller bias on the right while it will generate a large variance on the left of the cut-off point. 
What is important is that a case like this is not an unrealistic artifact but arises naturally in many empirical studies. 
For example, sharp contrasts in slopes are observed in Figures 1 and 2 in LM, Figures 12 and 14 in \citet{dm08}, Figures 3 and 5 in \citet{le08} and Figures 1 and 2 in \citet{bpl13}, among others.

A single bandwidth approach is familiar to empirical researchers in the applications of matching methods (\citet{ai11}) since the supports of covariates for treated and untreated individuals overlap and we wish to construct two comparable groups. 
This reasoning does not apply to the RD estimator since values of the
assignment variable never overlap due to the structure of the RD
design. 
Indeed, the observations to the left of the cut-off point is used to
approximate the left-limit point and the observations to the right of
the cut-off point is used to approximate the right-limit point so that
there is no reason the two bandwidths should be the same.
Although IK recognizes the appropriateness of choosing the bandwidths
separately on both sides of the discontinuity point, they proceed
to choose the same bandwidth to estimate two functions on both sides
of the discontinuity point to avoid the technical difficulty of the
cancelation of the first order bias terms in some cases (\citet[pp. 936--937]{ik12}).\footnote{We will discuss this issue in detail in section 2.}

The main contributions of this paper are to provide an approach to
resolve the technical difficulty of choosing the two different bandwidths
for the RD estimator, theoretically show the proposed method dominates
the IK method and other methods in terms of the asymptotic mean squared error (AMSE), and show that the theoretical advange materialize in empirically relevant sample sizes through an extensive simulation study. 
To the best of our knowledge, this paper is the first to consider choosing two bandwidths simultaneously in the RD context.\footnote{\citet{mp97} consider the optimal selection of two bandwidths to estimate the ratio of the first derivative of the density to the density itself. Since the optimal rates for the bandwidths for the numerator and the denominator differ in their case, their results do not apply in the present context.}

The paper is organized as follows. 
We first discuss the technical difficulty of the simultaneous selection of the bandwidths and describe how we define theoretically optimal bandwidths in the RD context. 
We next show that the proposed method dominates currently available methods in the AMSE sense. 
We then propose a feasible version of the bandwidth selection rule and show its asymptotic equivalence to the theoretically optimal bandwidths.
Finally we illustrate how the theoretical results realize in empirically relevant sample sizes via simulation experiments and an empirical example.
In the Appendix, we include the proofs of the main theorems.\footnote{Matlab and Stata codes to implement the proposed method are available as a part of the Supplementary Materials (or at one of the authors' webpage, \texttt{http://www3.grips.ac.jp/\~{}yarai/}).}

\section{Bandwidth Selection of The Sharp Regression Discontinuity Estimators}

\label{sec:rddestimation} 
For observation $i$ we denote potential outcomes with and without treatment by $Y_{i}(1)$ and $Y_{i}(0)$, respectively. 
Let $D_{i}$ be a binary variable ($0$ and $1$) indicating the treatment status. 
The observed outcome, $Y_{i}$, can be written as $Y_{i}=D_{i}Y_{i}(1)+(1-D_{i})Y_{i}(0)$. 
In the sharp RD setting we consider, the treatment status is determined solely by the assignment variable, denoted by $X_{i}$: $D_{i}=\mathbb{I}\{X_{i}\geq c\}$
where $c$ is a known cut-off point and $\mathbb{I}\{A\}$ takes value $1$ if $A$ holds and takes value $0$ if $A$ does not hold. 
Throughout the paper, we assume that $(Y_{1},X_{1})$, $\ldots$ , $(Y_{n},X_{n})$ are independent and identically distributed observations and $X_{i}$ has the Lebesgue density $f$.

Define $m_{1}(x)=E(Y_{i}(1)|X_{i}=x)=E(Y_{i}|X_{i}=x)$ for $x\geq c$ and $m_{0}(x)=E(Y_{i}(0)|X_{i}=x)=E(Y_{i}|X_{i}=x)$ for $x<c$. 
Suppose that the limits $\lim_{x\to c+}m_{1}(x)$ and $\lim_{x\to c-}m_{0}(x)$ exist where $x\to c+$ and $x\to c-$ mean taking the limits from
the right and left, respectively. Denote $\lim_{x\to c+}m_{1}(x)$ and $\lim_{x\to c-}m_{0}(x)$ by $m_{1}(c)$ and $m_{0}(c)$, respectively.
Then the average treatment effect at the cut-off point is given by $\tau(c)=m_{1}(c)-m_{0}(c)$ and $\tau(c)$ is the parameter of interest in the sharp RD design.

Estimation of $\tau(c)$ requires to estimate two functions, $m_{1}(c)$ and $m_{0}(c)$. The nonparametric estimators that we consider are LLR estimators proposed by \citet{st77} and investigated by \citet{fa92}.
For estimating these limits, the LLR is particularly attractive because it exhibits the automatic boundary adaptive property (\citet{fa92,fa93},
\citet{htv01}, and \citet{por03}). 
The LLR estimator for $m_{1}(c)$ is given by $\hat{\alpha}_{h_{1}}(c)$, where 
\begin{align*}
\left(\hat{\alpha}_{h_{1}}(c),\hat{\beta}_{h_{1}}(c)\right)=\arg\min_{\alpha,\beta}\sum_{i=1}^{n}\left\{ Y_{i}-\alpha-\beta(X_{i}-c)\right\} ^{2}K\left(\frac{X_{i}-c}{h_{1}}\right)\mathbb{I}\{X_{i}\geq c\},
\end{align*}
where $K(\cdot)$ is a kernel function and $h_{1}$ is a bandwidth.
A standard choice of the kernel function for the RD estimators is the triangular kernel given by $K(u)=(1-|u|)\mathbb{I}\{|u|<1\}$ because of its MSE and minimax optimality (\citet{cfm97}). 
The LLR estimator for $m_{0}(c)$, $\hat{\alpha}_{h_{0}}(c)$, can be obtained in the same manner. Denote $\hat{\alpha}_{h_{1}}(c)$ and $\hat{\alpha}_{h_{0}}(c)$ by $\hat{m}_{1}(c)$ and $\hat{m}_{0}(c)$, respectively. Then $\tau(c)$ is estimated by $\hat{m}_{1}(c)-\hat{m}_{0}(c)$.

\subsection{The AMSE for The Regression Discontinuity Estimators}

In this paper, we propose a simultaneous selection method for two distinct bandwidths, $h_{1}$ and $h_{0}$, based on an AMSE. 
This is also the standard approach in the literature.\footnote{As IK emphasize, the bandwidth selection problem in the context of
the RD setting is how to choose local bandwidths rather than global bandwidths.  Thus, bandwidth selection based on either the asymptotic mean ``integrated'' squared errors or the cross-validation criterion can never be optimal.}

The conditional MSE of the RD estimators given the assignment variable,
$X$, is defined by 
\[
MSE_{n}(h)=E\Bigl[\left.\bigl\{\left[\hat{m}_{1}(c)-\hat{m}_{0}(c)\right]-\left[m_{1}(c)-m_{0}(c)\right]\bigr\}^{2}\right|X\Bigr].
\]
where $X=(X_{1},X_{2},\ldots,X_{n})'$.\footnote{Throughout the paper, we use ``$h$'' without a subscript to denote a combination of $h_{1}$ and $h_{0}$; e.g., $MSE_{n}(h_{1},h_{0})$ is written as $MSE_{n}(h)$. We assume that $X$ is such that this
conditional MSE is well defined.} 
A standard approach is to obtain the AMSE, ignoring higher-order terms, and to choose the bandwidths that minimize it. 
To do so, we proceed under the following assumptions. (The integral sign $\int$ refers to an integral over the range $(-\infty,\infty)$ unless stated otherwise.) 
\begin{assumption}\label{assumption:kernel} $K(\cdot):\mathbb{R}\to\mathbb{R}$
is a symmetric second-order kernel function that is continuous with
compact support; i.e., $K$ satisfies the following: $\int K(u)du=1$,
$\int uK(u)du=0$, and $\int u^{2}K(u)du\ne0$. \end{assumption}

\begin{assumption}\label{assumption:bandwidth} 
The positive sequence of bandwidths is such that $h_{j}\to0$ and $nh_{j}\to\infty$ as
$n\to\infty$ for $j=0,1$. \end{assumption} 
Assumptions \ref{assumption:kernel} and \ref{assumption:bandwidth} are standard in the literature of regression function estimation.

Let ${\cal D}$ be an open set in $\mathbb{R}$, $k$ be a nonnegative integer, ${\cal C}_{k}$ be the family of $k$ times continuously differentiable functions on ${\cal D}$ and $g^{(k)}(\cdot)$ be the $k$th derivative of $g(\cdot)\in{\cal C}_{k}$. 
Let ${\cal G}_{k}({\cal D})$ be the collection of functions $g$ such that $g\in{\cal C}_{k}$ and 
\[
\left|g^{(k)}(x)-g^{(k)}(y)\right|\le M_{k}\left|x-y\right|^{\alpha},\quad x,y,z\in{\cal D},
\]
for some positive $M_{k}$ and some $\alpha$ such that $0<\alpha\le1$.

Let $\sigma_{1}^{2}(x)$ and $\sigma_{0}^{2}(x)$ denote the conditional variance of $Y_{1}$ and $Y_{0}$ given $X_{i}=x$, respectively and let $\sigma_{1}^{2}(c)=\lim_{x\to c+}\sigma_{1}^{2}(x)$, $\sigma_{0}^{2}(c)=\lim_{x\to c-}\sigma_{0}^{2}(x)$, $m_{1}^{(2)}(c)=\lim_{x\to c+}m_{1}^{(2)}(x)$, $m_{0}^{(2)}(c)=\lim_{x\to c-}m_{0}^{(2)}(x)$, $m_{1}^{(3)}(c)=\lim_{x\to c+}m_{1}^{(3)}(x)$, $m_{0}^{(3)}(c)=\lim_{x\to c-}m_{0}^{(3)}(x)$, $\mu_{j,0}=\int_{0}^{\infty}u^{j}K(u)du$ and $\nu_{j,0}=\int_{0}^{\infty}u^{j}K^{2}(u)du$ for nonnegative integer $j$.

\begin{assumption}\label{assumption:LLRBdensity} 
The Lebesgue density of $X_{i}$, denoted $f$, is an element of ${\cal G}_{1}({\cal D})$ where ${\cal D}$ is an open neighborhood of $c$ and is bounded above and strictly positive on ${\cal D}$ \end{assumption} 
\begin{assumption}\label{assumption:LLRBfunction}
Let $\delta$ be some positive constant. 
The conditional mean function $m_{1}$ and the conditional variance function $\sigma_{1}^{2}$ are elements of ${\cal G}_{3}({\cal D}_{1})$ and ${\cal G}_{0}({\cal D}_{1})$, respectively, where ${\cal D}_{1}$ is a one-sided open neighborhood of $c$, $(c,c+\delta)$, and $m_{1}(c)$, $m_{1}^{(2)}(c)$, $m_{1}^{(3)}(c)$ and $\sigma_{1}^{2}(c)$ exist and are bounded above and strictly positive. 
Analogous conditions hold for $m_{0}$ and $\sigma_{0}^{2}$ on ${\cal D}_{0}$, where ${\cal D}_{0}$ is a one-sided open neighborhood of $c$, $(c-\delta,c)$. \end{assumption}

Under Assumptions \ref{assumption:kernel}, \ref{assumption:bandwidth}, \ref{assumption:LLRBdensity} and \ref{assumption:LLRBfunction}, we can easily generalize the result obtained by \citet{fg92}:\footnote{The conditions on the first derivative of $f$ and the third derivatives of $m_{1}$ and $m_{0}$, described in Assumptions \ref{assumption:LLRBdensity} and \ref{assumption:LLRBfunction}, are not necessary to obtain the result (\ref{eq:mse}). They are stated for later use.} 
\begin{align}\label{eq:mse}
MSE_{n}(h) & =\left\{ \frac{b_{1}}{2}\left[m_{1}^{(2)}(c)h_{1}^{2}-m_{0}^{(2)}(c)h_{0}^{2}\right]\right\} ^{2}+\frac{v}{nf(c)}\left\{ \frac{\sigma_{1}^{2}(c)}{h_{1}}+\frac{\sigma_{0}^{2}(c)}{h_{0}}\right\} \nonumber \\
 & \hspace{5mm}+o\left(h_{1}^{4}+h_{1}^{2}h_{0}^{2}+h_{0}^{4}+\frac{1}{nh_{1}}+\frac{1}{nh_{0}}\right),
\end{align}
where 
\[
b_{1}=\frac{\mu_{2,0}^{2}-\mu_{1,0}\mu_{3,0}}{\mu_{0,0}\mu_{2,0}-\mu_{1,0}^{2}},\quad\mbox{and}\quad v=\frac{\mu_{2,0}^{2}\nu_{0,0}-2\mu_{1,0}\mu_{2,0}\nu_{1,0}+\mu_{1,0}^{2}\nu_{2,0}}{(\mu_{0,0}\mu_{2,0}-\mu_{1,0}^{2})^{2}}.
\]
This suggests that we choose the bandwidths to minimize the following AMSE: 
\begin{equation}
AMSE_{n}(h)=\left\{ \frac{b_{1}}{2}\left[m_{1}^{(2)}(c)h_{1}^{2}-m_{0}^{(2)}(c)h_{0}^{2}\right]\right\} ^{2}+\frac{v}{nf(c)}\left\{ \frac{\sigma_{1}^{2}(c)}{h_{1}}+\frac{\sigma_{0}^{2}(c)}{h_{0}}\right\} .\label{eq:standardAMSE}
\end{equation}
However, this procedure can fail. To see why, let $h_{1}$, $h_{0}\in H$, where $H=(0,\infty)$, and consider the case in which $m_{1}^{(2)}(c)m_{0}^{(2)}(c)>0$.
Now choose $h_{0}=[m_{1}^{(2)}(c)/m_{0}^{(2)}(c)]^{1/2}h_{1}$. 
Then, we have 
\begin{align*}
AMSE_{n}(h) & =\frac{v}{nh_{1}f(c)}\left\{ \sigma_{1}^{2}(c)+\sigma_{0}^{2}(c)\left[\frac{m_{0}^{(2)}(c)}{m_{1}^{(2)}(c)}\right]^{1/2}\right\} .
\end{align*}
This implies that the bias component can be removed completely from the AMSE by choosing a specific ratio of bandwidths and the AMSE can be made arbitrarily small by choosing a sufficiently large $h_{1}$.

One reason for this nonstandard behavior is that the AMSE given in (\ref{eq:standardAMSE}) does not account for higher-order terms.
If we account for the higher-order terms for the bias component, they should punish the act of choosing large values for bandwidths. 
In what follows, we show that simply incorporating the second-order bias
term into the AMSE does not resolve the problem. After demonstrating
this, we propose an alternative objective function that defines the target bandwidths.

The next lemma presents the MSE with a second-order bias term by generalizing the higher-order approximation of \citet{fghh96}.\footnote{\citet{fghh96} show the higher-order approximation of the MSE for interior points of the support of $X$. Lemma \ref{lemma:LLRMSEb} presents the analogous result for a boundary point.
A proof of Lemma 1 is provided in the Supplementary Material.} 
\begin{lemma}\label{lemma:LLRMSEb} 
Suppose Assumptions \ref{assumption:kernel}--\ref{assumption:LLRBfunction} hold. 
Then, it follows that 
\begin{align*}
MSE_{n}(h) & =\left\{ \frac{b_{1}}{2}\left[m_{1}^{(2)}(c)h_{1}^{2}-m_{0}^{(2)}(c)h_{0}^{2}\right]+\Bigl[b_{2,1}(c)h_{1}^{3}-b_{2,0}(c)h_{0}^{3}\Bigr]+o\left(h_{1}^{3}+h_{0}^{3}\right)\right\} ^{2}\\
 & \hspace{5mm}+\frac{v}{nf(c)}\left\{ \frac{\sigma_{1}^{2}(c)}{h_{1}}+\frac{\sigma_{0}^{2}(c)}{h_{0}}\right\} +o\left(\frac{1}{nh_{1}}+\frac{1}{nh_{0}}\right),
\end{align*}
where, for $j=0,1$, 
\begin{align*}
 & b_{2,j}(c)=(-1)^{j+1}\left\{ \xi_{1}\left[\frac{m_{j}^{(2)}(c)}{2}\frac{f^{(1)}(c)}{f(c)}+\frac{m_{j}^{(3)}(c)}{6}\right]-\xi_{2}\frac{m_{j}^{(2)}(c)}{2}\frac{f^{(1)}(c)}{f(c)}\right\} \\
 & \xi_{1}=\frac{\mu_{2,0}\mu_{3,0}-\mu_{1,0}\mu_{4,0}}{\mu_{0,0}\mu_{2,0}-\mu_{1,0}^{2}},\quad{\mbox{a}nd}\quad\xi_{2}=\frac{(\mu_{2,0}^{2}-\mu_{1,0}\mu_{3,0})\left(\mu_{0,0}\mu_{3,0}-\mu_{1,0}\mu_{2,0}\right)}{(\mu_{0,0}\mu_{2,0}-\mu_{1,0}^{2})^{2}}.
\end{align*}
\end{lemma}

Given the expression of Lemma \ref{lemma:LLRMSEb}, one might be tempted to proceed with an AMSE including the second-order bias term: 
\begin{multline}
AMSE2_{n}\equiv\left\{ \frac{b_{1}}{2}\left[m_{1}^{(2)}(c)h_{1}^{2}-m_{0}^{(2)}(c)h_{0}^{2}\right]+\Bigl[b_{2,1}(c)h_{1}^{3}-b_{2,0}(c)h_{0}^{3}\Bigr]\right\} ^{2}\\
+\frac{v}{nf(c)}\left\{ \frac{\sigma_{1}^{2}(c)}{h_{1}}+\frac{\sigma_{0}^{2}(c)}{h_{0}}\right\} \label{amse2}
\end{multline}
We show that the minimization problem is not well-defined when $m_{1}^{(2)}(c)m_{0}^{(2)}(c)>0$.
In particular, we show that one can make the order of the bias term $O(h_{1}^{k+3})$, for an arbitrary nonnegative interger $k$, by choosing $h_{0}^{2}=C(h_{1},k)h_{1}^{2}$ and $C(h_{1},k)=C_{0}+C_{1}h_{1}+C_{2}h_{1}^{2}+C_{3}h_{1}^{3}+\text{\ensuremath{\ldots}}+C_{k}h_{1}^{k}$ for some constants $C_{0}$, $C_{1}$, $\ldots$, $C_{k}$ when the sign of the product of the second derivatives is positive. 
Given that bandwidths are necessarily positive, we must have $C_{0}>0$, although we allow $C_{1}$, $C_{2}$, $\ldots$, $C_{k}$ to be negative. 
For sufficiently large $n$ and for any $k$, we always have $C(h_{1},k)>0$ given $C_{0}>0$ and we assume this without loss of generality.

To gain insight, consider choosing $C(h_{1},1)=C_{0}+C_{1}h_{1}$, where $C_{0}=m_{1}^{(2)}(c)/m_{0}^{(2)}(c)$. In this case, the sum of the first- and second-order bias terms is 
\begin{align*}
 & \frac{b_{1}}{2}\left[m_{1}^{(2)}(c)-C(h_{1},1)m_{0}^{(2)}(c)\right]h_{1}^{2}+\left[b_{2,1}(c)-C(h_{1},1)^{3/2}b_{2,0}(c)\right]h_{1}^{3}\\
 & \quad=\left\{ -\frac{b_{1}}{2}C_{1}m_{0}^{(2)}(c)+b_{2,1}(c)-C_{0}^{3/2}b_{2,0}(c)\right\} h_{1}^{3}+O(h_{1}^{4}).
\end{align*}
By choosing $C_{1}=2\left[b_{2,1}(c)-C_{0}^{3/2}b_{2,0}(c)\right]\Big/\left[b_{1}m_{0}^{(2)}(c)\right]$, one can make the order of bias $O(h_{1}^{4})$. 
Next, consider $C(h_{1},2)=C_{0}+C_{1}h_{1}+C_{2}h_{1}^{2}$, where $C_{0}$ and $C_{1}$ are as determined above. 
In this case,
\begin{align*}
 & \frac{b_{1}}{2}\left[m_{1}^{(2)}(c)-C(h_{1},2)m_{0}^{(2)}(c)\right]h_{1}^{2}+\left[b_{2,1}(c)-C(h_{1},2)^{3/2}b_{2,0}(c)\right]h_{1}^{3}\\
 & \quad=-\left\{ b_{1}C_{2}m_{0}^{(2)}(c)+3C_{0}^{1/2}C_{1}b_{2,0}(c)\right\} h_{1}^{4}/2+O(h_{1}^{5}).
\end{align*}
Hence, by choosing $C_{2}=-3C_{0}^{1/2}C_{1}b_{2,0}(c)/[b_{1}m_{0}^{(2)}(c)]$, one can make the order of bias term $O(h_{1}^{5})$. 
Similar arguments can be formulated for arbitrary $k$: the discussion above is summarized in the following lemma. 
\begin{lemma} 
Suppose Assumptions \ref{assumption:kernel}--\ref{assumption:LLRBfunction} hold. 
Also suppose $m_{1}^{(2)}(c)m_{0}^{(2)}(c)>0$. 
Then there exist a combination of $h_{1}$ and $h_{0}$ such that the AMSE including the second-order bias term defined in (\ref{amse2}) becomes 
\[
\frac{v}{nh_{1}f(c)}\left\{ \sigma_{1}^{2}(c)+\sigma_{0}^{2}(c)\left[\frac{m_{1}^{(2)}(c)}{m_{0}^{(2)}(c)}\right]^{1/2}\right\} +O\left(h_{1}^{k+3}\right).
\]
for an arbitrary nonnegative integer $k$. \end{lemma}

This implies non-existence of the optimal solution because one can choose $h_{1}$ arbitrarily close to $1$ and choose $k$ to diverge as $n$ becomes large. 
It is straightforward to generalize this discussion to the case of the AMSE with higher-order bias terms.\footnote{In the present approach, we consider choosing the bandwidths for the LLR estimator. In the literature of regression function estimation, it is common to employ local polynomial regression (LPR) of second-order when the conditional mean function is three times continuously differentiable because it is known to reduce bias (see, e.g., \citealp{fa92}). However, we have two reasons for confining our attention to the LLR. First, as shown later, we can achieve the same bias reduction with the LLR when the sign of the product of the second derivatives is positive. When the sign is negative, the existence of the third derivatives becomes unnecessary. Second, even when we use a higher order LPR, we end up with an analogous problem. For example, the first-order bias term is removed by using the second order LPR, but when the signs of $b_{2,1}(c)$ and $b_{2,0}(c)$ are the same, the second-order bias term can be eliminated by using an appropriate choice of bandwidths.}

\subsection{AFO Bandwidths}

\label{sec:AFOB} 
In order to overcome the difficulty just discussed, we propose a new
optimality criterion termed ``asymptotic first-order optimality'' (AFO).

When $m_{1}^{(2)}(c)m_{0}^{(2)}(c)<0$, there is no problem with the standard AMSE given by equation (\ref{eq:standardAMSE}). 
Hence we use this criterion.  When $m_{1}^{(2)}(c)m_{0}^{(2)}(c)>0$, we
choose $h_{0}^{2}=C_{0}h_{1}^{2}$ with
$C_{0}=m_{1}^{(2)}(c)/m_{0}^{(2)}(c)$, so that the first order
bias component vanishes and use the trade-off between the second-order
bias term and the asymptotic variance term to choose the bandwidths.
This amounts to using the AMSE with the second-order bias and the
asymptotic variance terms under the
restriction that the first-order bias term vanishes as the criterion.
The above discussion is formalized in the following definition. The resulting bandwidths are termed ``AFO bandwidths.''\footnote{We note that the asymptotically higher-order optimal bandwidths can be proposed in the same manner under an additional smoothness condition.
We do not pursue this direction further in this paper because of implementation difficulty. 
More detailed discussions are provided in \citet{ai13}.}

\begin{define}\label{def:FirstOptimalRegB} 
The AFO bandwidths for the RD estimator {\rm minimize the AMSE defined by 
\[
{AMSE}_{1n}(h)=\left\{ \frac{b_{1}}{2}\left[m_{1}^{(2)}(c)h_{1}^{2}-m_{0}^{(2)}(c)h_{0}^{2}\right]\right\} ^{2}+\frac{v}{nf(c)}\left\{ \frac{\sigma_{1}^{2}(c)}{h_{1}}+\frac{\sigma_{0}^{2}(c)}{h_{0}}\right\} .
\]
when $m_{1}^{(2)}(c)m_{0}^{(2)}(c)<0$. Their explicit expressions are given by $h_{1}^{*}=\theta^{*}n^{-1/5}$ and $h_{0}^{*}=\lambda^{*}h_{1}^{*}$, where 
\begin{equation}
\theta^{*}=\left\{ \frac{v\sigma_{1}^{2}(c)}{b_{1}^{2}f(c)m_{1}^{(2)}(c)\left[m_{1}^{(2)}(c)-{\lambda^{*}}^{2}m_{0}^{(2)}(c)\right]}\right\} ^{1/5}\quad\mbox{and}\quad\lambda^{*}=\left\{ -\frac{\sigma_{0}^{2}(c)m_{1}^{(2)}(c)}{\sigma_{1}^{2}(c)m_{0}^{(2)}(c)}\right\} ^{1/3}.\label{eq:starparameter}
\end{equation}
When} $m_{1}^{(2)}(c)m_{0}^{(2)}(c)>0$, the AFO bandwidths for the RD estimator {\rm minimize the AMSE defined by 
\[
AMSE_{2n}(h)=\Bigl\{ b_{2,1}(c)h_{1}^{3}-b_{2,0}(c)h_{0}^{3}\Bigr\}^{2}+\frac{v}{nf(c)}\left\{ \frac{\sigma_{1}^{2}(c)}{h_{1}}+\frac{\sigma_{0}^{2}(c)}{h_{0}}\right\} 
\]
subject to the restriction $m_{1}^{(2)}(c)h_{1}^{2}-m_{0}^{(2)}(c)h_{0}^{2}=0$ under the assumption of $b_{2,1}(c)-\{m_{1}^{(2)}(c)/m_{0}^{(2)}(c)\}^{3/2}b_{2,0}(c)\ne0$.
Their explicit expressions are given by $h_{1}^{**}=\theta^{**}n^{-1/7}$ and $h_{0}^{**}=\lambda^{**}h_{1}^{**}$, where 
\begin{equation}
\theta^{**}=\left\{ \frac{v\left[\sigma_{1}^{2}(c)+\sigma_{0}^{2}(c)/\lambda^{**}\right]}{6f(c)\left[b_{2,1}(c)-{\lambda^{**}}^{3}b_{2,0}(c)\right]^{2}}\right\} ^{1/7}\quad\mbox{and}\quad\lambda^{**}=\left\{ \frac{m_{1}^{(2)}(c)}{m_{0}^{(2)}(c)}\right\} ^{1/2}.\label{eq:starparameter2}
\end{equation}
}
 \end{define} \vspace{1mm}
Definition \ref{def:FirstOptimalRegB} is stated assuming that the first- and the second-order bias terms do not vanish simultaneously, i.e., $b_{2,1}(c)-\{m_{1}^{(2)}(c)/m_{0}^{(2)}(c)\}^{3/2}b_{2,0}(c)\ne0$ when $m_{1}^{(2)}(c)m_{0}^{(2)}(c)>0$.\footnote{Uniqueness of the AFO bandwidths in each case is verified in \citet{ai13b}.
Definition \ref{def:FirstOptimalRegB} can be generalized to cover the excluded case in a straightforward manner if we are willing to assume the existence of the fourth derivatives. 
This case corresponds to the situation in which the first- and the second-order bias terms can be removed simultaneously by choosing appropriate bandwidths and the third-order bias term works as a penalty for large bandwidths. 
Another excluded case in Definition \ref{def:FirstOptimalRegB} is when $m_{1}^{(2)}(c)m_{0}^{(2)}(c)=0$. It is also possible to extend the idea of the AFO bandwidths when both $m_{1}^{(2)}(c)=0$ and $m_{0}^{(2)}(c)=0$ hold. 
This generalization can be carried out by replacing the role of the first- and the second-order bias terms by the second- and the third order bias terms.}


The proposed approach based on the AFO bandwidths asymptotically dominates the existing approaches in the AMSE, irrespective of the values of the second derivatives.
To see this, first note that when the product of the second derivatives is positive, the AMSE based on the AFO bandwidths is of order $n^{-6/7}$ whereas the AMSE based on the optimal bandwidths chosen for each of the regression function separately (we refer to this bandwidths, Independent Bandwidths (IND)) is of order $n^{-4/5}$.\footnote{The independent selection chooses the bandwidths on the left and the right of the cut-off optimally for each function without paying attention to the relationship between the two functions. 
The IND bandwidths based on the AMSE criterion are given by 
\[
\check{h}_{1}=\left\{ \frac{v\sigma_{1}^{2}(c)}{b_{1}^{2}f(c)\left[m_{1}^{(2)}(c)\right]^{2}}\right\} ^{1/5}n^{-1/5}\quad\mbox{and}\quad\check{h}_{0}=\left\{ \frac{v\sigma_{0}^{2}(c)}{b_{1}^{2}f(c)\left[m_{0}^{(2)}(c)\right]^{2}}\right\} ^{1/5}n^{-1/5}.
\]
}
The same is true for the IK bandwidth unless the two second derivatives are exactly the same. 
Thus when the product of the second derivatives is positive, AFO bandwidths are more efficient than either the IK bandwidth or the IND bandwidths in the sense that the AMSE have a faster rate of convergence.

The only exception to this rule is when the second derivatives are the same. 
In this case, the IK bandwidth is 
\[
h_{IK} = \theta_{IK}  n^{-1/7}
\]
where
\[
\theta_{IK} = C_{IK} \left(\frac{\sigma_{1}^{2}(c) + \sigma_{0}^{2}(c)}{ [\sigma_{1}^{2}(c)]^{2/7} \{p_{1} [m_{1}^{(3)}(c)]^{2}\}^{5/7} + [\sigma_{0}^{2}(c)]^{2/7}\{p_{0} [m_{0}^{(3)}(c)]^{2}\}^{5/7} } \right)^{1/5},
\]
$C_{IK} = [v/(2160\cdot 3.56^{5} \cdot b_{1}^{2} [f(c)]^{5/7}) ]^{1/5}$, $p_{1}=\int_{c}^{\infty}f(x)dx$ and $p_{0}=\int_{-\infty}^{c}f(x)dx$.\footnote{The derivation of $\theta_{IK}$ is provided in the Supplementary Material.}
Although this bandwidth is of the same order with the AFO bandwidths, it is not determined by minimizing the AMSE. 
In fact the ratio of the AMSE up to the second-order bias term based on the AFO bandwidths to that of the IK bandwidth converges to 
\[
\frac{1}{{\displaystyle {\frac{1}{7}}\gamma^{6}+{\displaystyle {\frac{6}{7}\frac{1}{\gamma}}}}},
\]
where $\gamma=\theta_{IK}/\theta_{AFO}$ and $\theta_{AFO}$ equals
$\theta^{**}$ in equation (5) when $\lambda^{**}=1$.\footnote{To see why
the ratios of the AMSEs converges to the specified limit, note that the ratio of the AMSEs is
\[
\frac{[b_{2,1}(c)-b_{2,0}(c)]^{2}\theta_{AFO}^{6}+{\displaystyle \frac{v [\sigma_{1}^{2}(c)+\sigma_{0}^{2}(c)]}{\theta_{AFO}f(c)}}}{[b_{2,1}(c)-b_{2,0}(c)]^{2}\theta_{IK}^{6}+{\displaystyle \frac{v [\sigma_{1}^{2}(c)+\sigma_{0}^{2}(c)]}{\theta_{IK}f(c)}}}.
\]
 Since the first order condition implies $v [\sigma_{1}^{2}(c)+\sigma_{0}^{2}(c)]/[\theta_{AFO} f(c))]=6 [b_{2,1}(c) - b_{2,0}(c)]^{2} \theta_{AFO}^{6}$, substituting this expression and after some simple calculations yield the result.} 
It is easy to show that the ratio is strictly less than one and equals one if and only if $\gamma=1$.  Since the $\theta_{AFO}$ depends on the second derivatives but $\theta_{IK}$ does not, the ratio can be much larger or smaller than one and hence ratio can be arbitrarily close to 0. 


When the sign of the product of the second derivatives is negative, the rates of convergence of the AMSEs corresponding to different bandwidth selection rules are the same. 
By construction, AFO bandwidths have the lowest AMSE.
The AMSEs corresponding to the AFO bandwidths, IK bandwidth, and IND bandwidths are, respectively 
\begin{align*}
AMSE(h^{*}) & =\frac{5}{4}n^{-4/5}C_{K}[m_{0}^{(2)}(c)]^{2/5}[\sigma_{0}^{2}(c)]^{4/5}][(\gamma_{1}\gamma_{2}^{2})^{1/3}+1]^{6/5},\\
AMSE(h_{IK}) & =\frac{5}{4}n^{-4/5}C_{K}[m_{0}^{(2)}(c)]^{2/5}[\sigma_{0}^{2}(c)]^{4/5}](\gamma_{1}+1)^{2/5}(\gamma_{2}+1)^{4/5},\text{ and}\\
AMSE(h_{IND}) & =\frac{5}{4}n^{-4/5}C_{K}[m_{0}^{(2)}(c)]^{2/5}[\sigma_{0}^{2}(c)]^{4/5}]((\gamma_{1}\gamma_{2}^{2})^{1/5}+1)^{2}((\gamma_{1}\gamma_{2}^{2})^{2/5}+1),
\end{align*}
where $\gamma_{1}=-m_{1}^{{(2)}}(c)/m_{0}^{{(2)}}(c)$$,$ $\gamma_{2}=\sigma_{1}^{2}(c)/\sigma_{0}^{2}(c)$,
and $C_{K}=[b_{1}v_{2}/f(c)^{2}]^{2/5}$.

\begin{landscape}
\begin{figure}[htbp]
  \begin{subfigure}[b]{0.33\linewidth}
    \centering
    \captionsetup{justification=centering}
    \includegraphics[width=1\linewidth]{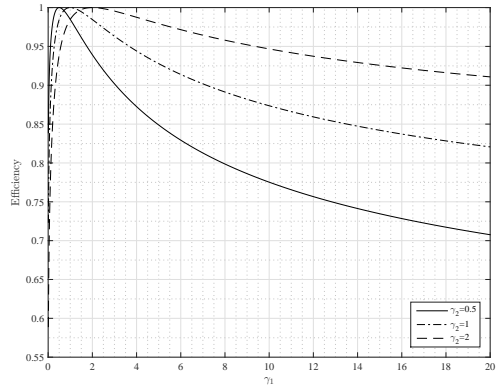} 
    \captionsetup{singlelinecheck=off}
    \caption{Efficiency as a function of $\gamma_{1}$ given $\gamma_{2}$}
    \label{figure:amse:a} 
    \vspace{4ex}
  \end{subfigure}
  \begin{subfigure}[b]{0.33\linewidth}
    \centering
    \captionsetup{justification=centering}
    \includegraphics[width=1\linewidth]{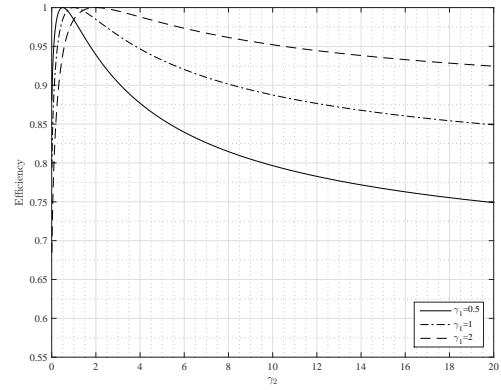} 
    \captionsetup{singlelinecheck=off}
    \caption{Efficiency as a function of $\gamma_{2}$ given $\gamma_{1}$}
    \label{figure:amse:b} 
    \vspace{4ex}
  \end{subfigure} 
  \begin{subfigure}[b]{0.33\linewidth}
    \centering
    \captionsetup{justification=centering}
    \includegraphics[width=1\linewidth]{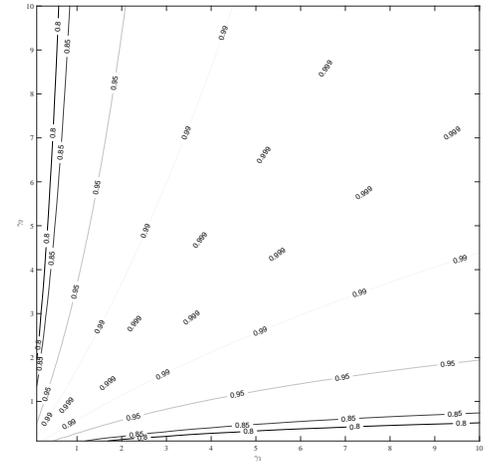} 
    \captionsetup{singlelinecheck=off}
    \caption{Contour as a function of $\gamma_{1}$ and $\gamma_{2}$}
    \label{figure:amse:c} 
    \vspace{4ex}
  \end{subfigure}
\caption{The ratio of the AMSEs, $AMSE(h^{*})/AMSE(h_{IK})$, as a function of $\gamma_{1}$ and $\gamma_{2}$. (a) The ratio of the AMSEs as a function of $\gamma_{1}$ given $\gamma_{2}$, (b) The ratio of the AMSEs as a function of $\gamma_{2}$ given $\gamma_{1}$, and (c) The contour of the ratio of the AMSEs, as a function of $\gamma_{1}$ and $\gamma_{2}$.}
\label{fig:complambda}
\end{figure}
\end{landscape}

Clearly the AMSE of the AFO relative to that of the IK depends only on  $\gamma_{1}$ and $\gamma_{2}$.
Efficiency as a function of $\gamma_{1}$ given $\gamma_{2}$ and that as a function of $\gamma_{2}$ given $\gamma_{1}$ are plotted in Figure \ref{fig:complambda}-(a) and Figure \ref{fig:complambda}-(b), and the contour of the ratio is depicted in Figure \ref{fig:complambda}-(c). 
It is straitforward to show that the maximum of the ratio
is 1 and attained if and only if $\gamma_{1}=\gamma_{2}$. 
We note that while the region on which the ratio is close to 1 is large, the ratio is less than 0.8 whenever $\gamma_{1}$ and $\gamma_{2}$ are rather different.

The AFO bandwidths have the advantage over the IND bandwidths. 
Again, clearly the AMSE of the AFO relative to that of the IND only depends on $\gamma_{1}$ and $\gamma_{2}$.
The ratio attains its minimum when $\gamma_{1}\gamma_{2}^{2}=1$ and the minimum value is $2^{6/5}/(12/5)\doteq0.957$.
It is an interesting finding that when the sign of the second derivatives differ, there is less than 5\% gain in efficiency by AFO over IND.

The AFO bandwidths improve the rate of convergence of AMSE when
the sign of the product of the second derivatives
is positive.  When the sign is negative, it is more efficient than
either the IK bandwidth or the IND bandwidths although the gain over IND
is less than 5\%.

\subsection{Feasible Automatic Bandwidth Choice}
\label{sec:feasible} 
The AFO bandwidths are clearly not feasible because they depend on unknown quantities related to $f(\cdot)$, $m_{1}$, and $m_{0}$.
An obvious plug-in version of the AFO bandwidths can be implemented by estimating these objects. 
Depending on the estimated sign of the product of the second derivatives, we can construct the plug-in version of the AFO bandwidths provided in Definition \ref{def:FirstOptimalRegB}.
We refer to these as ``the direct plug-in AFO bandwidths.''\footnote{The direct plug-in AFO bandwidths are defined by 
\begin{align*}
\hat{h}_{1}^{D} & =\hat{\theta}^{*}n^{-1/5}\mathbb{I}\{\hat{m}_{1}^{(2)}(c)\hat{m}_{0}^{(2)}(c)<0\}+\hat{\theta}^{**}n^{-1/7}\mathbb{I}\{\hat{m}_{1}^{(2)}(c)\hat{m}_{0}^{(2)}(c)\geq0\},\\
\hat{h}_{0}^{D} & =\hat{\theta}^{*}\hat{\lambda}^{*}n^{-1/5}\mathbb{I}\{\hat{m}_{1}^{(2)}(c)\hat{m}_{0}(c)<0\}+\hat{\theta}^{**}\hat{\lambda}^{**}n^{-1/7}\mathbb{I}\{\hat{m}_{1}^{(2)}(c)\hat{m}_{0}^{(2)}(c)\geq0\},
\end{align*}
where $\hat{\theta}^{*}$, $\hat{\lambda}^{*}$, $\hat{\theta}^{**}$ and $\hat{\lambda}^{**}$ are consistent estimators for $\theta^{*}$, $\lambda^{*}$, $\theta^{**}$ and $\lambda^{**}$ defined in (\ref{eq:starparameter}) and (\ref{eq:starparameter2}), respectively.} 
We can show that the direct plug-in AFO bandwidths are asymptotically as good as the AFO bandwidths in large samples. That is, we can prove that a version of Theorem 1 below also holds for the direct plug-in AFO bandwidths. 
However, our unreported simulation experiments show a poor performance of the direct plug-in AFO bandwidths under the designs described in Section \ref{sec:simulation} possibly because they misjudge the rate of the bandwidths whenever the sign is misjudged. 
Hence we do not pursue the direct plug-in approach further.

Instead, we propose an alternative procedure for choosing bandwidths that switch between two bandwidths more smoothly. To propose feasible bandwidths, we present a modified version of the AMSE (MMSE) defined by 
\begin{align*}
MMSE_{n}(h) & =\left\{ \frac{b_{1}}{2}\left[m_{1}^{(2)}(c)h_{1}^{2}-m_{0}^{(2)}(c)h_{0}^{2}\right]\right\} ^{2}+\Bigl\{ b_{2,1}(c)h_{1}^{3}-b_{2,0}(c)h_{0}^{3}\Bigr\}^{2}\\
 & \hspace{5mm}+\frac{v}{nf(x)}\left\{ \frac{\sigma_{1}^{2}(x)}{h_{1}}+\frac{\sigma_{0}^{2}(x)}{h_{0}}\right\} .
\end{align*}
A notable characteristic of the MMSE is that the bias component is represented by the sum of the squared first- and the second-order bias terms. 
Therefore, its bias component cannot be made arbitrarily small even when the sign is positive, unless $b_{2,1}(c)-\{m_{1}^{(2)}(c)/m_{0}^{(2)}(c)\}^{3/2}b_{2,0}(c) = 0$.
Thus, either term can penalize large bandwidths regardless of the sign so that the MMSE preserves the bias-variance trade-off in contrast to the AMSE with the second-order bias term. 
When $m_{1}^{(2)}(c)m_{0}^{(2)}(c)<0$, the square of the first-order bias term serves as the leading penalty and that of the second-order bias term becomes the second-order penalty.
When $m_{1}^{(2)}(c)m_{0}^{(2)}(c)>0$, the square of the second-order bias term works as the penalty and that of the first-order bias term becomes the linear restriction that shows up in the definition of the AFO bandwidths. 
In fact, the bandwidths that minimize the MMSE are asymptotically equivalent to the AFO bandwidths. 
This claim can be proved rigorously as a special case of the following theorem.

We propose a feasible bandwidth selection method based on the MMSE.
The proposed method for bandwidth selection can be considered as a generalization of the traditional plug-in method (see, e.g., \citealp[Section 3.6]{wj94}).
Consider the following plug-in version of the MMSE denoted by $MMSE^{p}$:
\begin{align}
MMSE_{n}^{p}(h) & =\left\{ \frac{b_{1}}{2}\left[\hat{m}_{1}^{(2)}(c)h_{1}^{2}-\hat{m}_{0}^{(2)}(c)h_{0}^{2}\right]\right\} ^{2}+\Bigl\{\hat{b}_{2,1}(c)h_{1}^{3}-\hat{b}_{2,0}(c)h_{0}^{3}\Bigr\}^{2}\nonumber \\
 & \hspace{5mm}+\frac{v}{n\hat{f}(c)}\left\{ \frac{\hat{\sigma}_{1}^{2}(c)}{h_{1}}+\frac{\hat{\sigma}_{0}^{2}(c)}{h_{0}}\right\} ,\label{eq:LLRMMSEhatB}
\end{align}
where $\hat{m}_{j}^{(2)}(c)$, $\hat{b}_{2,j}(c)$, $\hat{\sigma}_{j}^{2}(c)$ and $\hat{f}(c)$ are consistent estimators of $m_{j}^{(2)}(c)$, $b_{2,j}(c)$, $\sigma_{j}^{2}(c)$ and $f(x)$ for $j=0,1$, respectively.
Let $(\hat{h}_{1},\hat{h}_{0})$ be a combination of bandwidths that minimizes the $MMSE^{p}$ given in (\ref{eq:LLRMMSEhatB}) and $\hat{h}$
denote $(\hat{h}_{1},\hat{h}_{0})$.\footnote{It is also possible to construct another version of the $MMSE^{p}$ based on the finite sample approximations discussed by \citet[Section 4.3]{fg96}.
We do not pursue this direction because it is computationally intensive for large sample and an unreported simulation produced the almost same result as that based on the $MMSE^{p}$ given in (\ref{eq:LLRMMSEhatB}).} 
In the next theorem, we show that $(\hat{h}_{1},\hat{h}_{0})$ is asymptotically as good as the AFO bandwidths in the sense of \citet{ha83} (see equation (2.2) of \citealp{ha83}).

\begin{theorem}\label{theorem:LLRpluginB} Suppose that the conditions stated in Lemma \ref{lemma:LLRMSEb} hold. 
Assume further that $\hat{m}_{j}^{(2)}(c)$, $\hat{b}_{2,j}(c)$, $\hat{f}(c)$ and $\hat{\sigma}_{j}^{2}(c)$ satisfy $\hat{m}_{j}^{(2)}(c)\plim m_{j}^{(2)}(c)$, $\hat{b}_{2,j}(c)\plim b_{2,j}(c)$, $\hat{f}(c)\plim f(c)$ and $\hat{\sigma}_{j}^{2}(c)\plim\sigma_{j}^{2}(c)$ for $j=0,1$, respectively. 
Then, the following hold. 
\begin{itemize}
\item[{\rm (i)}] When $m_{1}^{(2)}(c)m_{0}^{(2)}(c)<0$, 
\[
\frac{\hat{h}_{1}}{h_{1}^{*}}\plim1,\quad\frac{\hat{h}_{0}}{h_{0}^{*}}\plim1,\quad\mbox{and}\quad\frac{MMSE_{n}^{p}(\hat{h})}{MSE_{n}(h^{*})}\plim1.
\]

\item[{\rm (ii)}] When $m_{1}^{(2)}(c)m_{0}^{(2)}(c)>0$ and $b_{2,1}(c)-\{m_{1}^{(2)}(c)/m_{0}^{(2)}(c)\}^{3/2}b_{2,0}(c)\ne0$
\[
\frac{\hat{h}_{1}}{h_{1}^{**}}\plim1,\quad\frac{\hat{h}_{0}}{h_{0}^{**}}\plim1,\quad\mbox{and}\quad\frac{MMSE_{n}^{p}(\hat{h})}{MSE_{n}(h^{**})}\plim1.
\]
\end{itemize}
\end{theorem} \vspace{1mm}

The first part of Theorem \ref{theorem:LLRpluginB} (i) and (ii) implies that the bandwidths that minimize the MMSE are asymptotically equivalent to the AFO bandwidths regardless of the sign of the product. 
The second part shows that the minimized value of the plug-in version of the MMSE is asymptotically the same as the MSE evaluated at the AFO bandwidths.
These two findings show that the bandwidths that minimize the MMSE possess the desired asymptotic properties. 
These findings also justify the use of the MMSE as a criterion function.\footnote{Theorem \ref{theorem:LLRpluginB} requires pilot estimates for $m_{j}^{(2)}(c)$, $b_{2,j}(c)$, $f(c)$ and $\sigma_{j}^{2}(c)$ for $j=0,1$. 
A detailed procedure about how to obtain the pilot estimates is given in the Supplemental Material.}

\section{Simulation}

\label{sec:simulation} 
To investigate the finite sample performance of the proposed method, we conducted simulation experiments. 
Our simulation experiments demonstrate that the theoretical advantages of the feasible AFO bandwidths have over the existing bandwidth selection rules, such as the IK bandwidth and the IND bandwidths, realize in the sample sizes relevant for empirical studies in general, and especially so for the simulation designs taken directly from empirical studies.

\subsection{Simulation Designs}

We consider four designs.  Designs 1--3 are the ones used for simulation experiments in the present context by IK and \citet{cct14} (hereafter CCT).  
Designs 1 and 2 are motivated by the empirical studies of \citet{le08} and \citet{lm07}, respectively.  Design 4 mimics the situation considered by \citet{lm07} where they investigate the effect of Head Start assistance on Head Start spending in 1968. 
This design corresponds to Panel A of Figure II in \citet[p. 176]{lm07}.\footnote{We followed IK and CCT to obtain the functional form. 
We fit the fifth-order global polynomial with different coefficients for the right and the left of the cut-off point after rescaling.}

\begin{figure}[h]
  \begin{subfigure}[b]{0.5\linewidth}
    \centering
    \captionsetup{justification=centering}
    \includegraphics[width=1\linewidth]{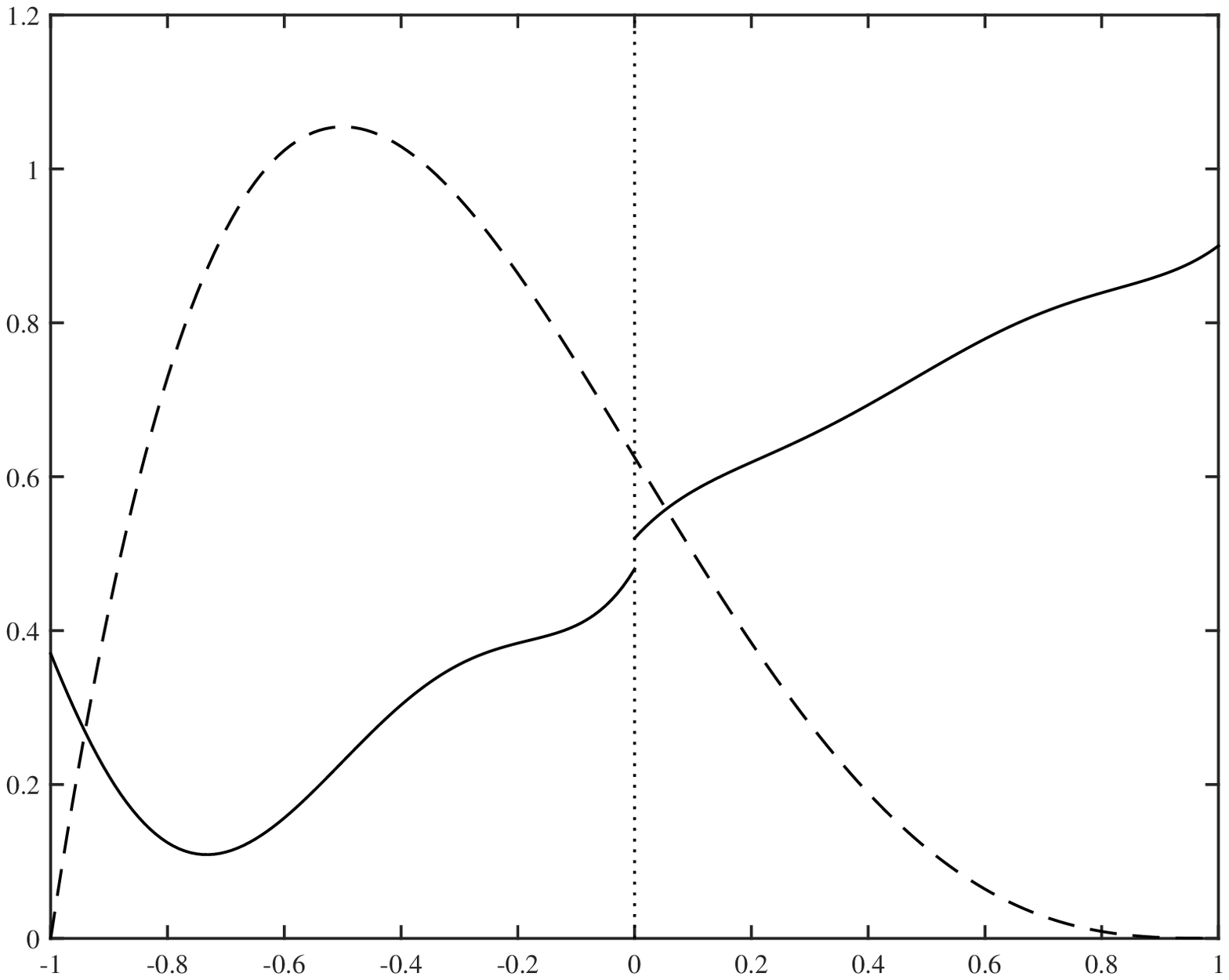} 
    \captionsetup{singlelinecheck=off}
    \caption[.]{ Design 1. Lee (2008) Data\\
    \hspace*{4mm} (Design 1 of IK and CCT)
    {\scriptsize
     \begin{align*}
    m_1(x) &= 0.52 + 0.84x - 3.00x^2 + 7.99x^3 - 9.01x^4 + 3.56x^5\\
    m_{0}(x) &= 0.48 + 1.27x + 7.18x^2 + 20.21x^3 + 21.54x^4 +7.33x^5 
   \end{align*}}} 
    \label{figure:dgp:a} 
    \vspace{4ex}
  \end{subfigure}
  \hspace{5mm}
  \begin{subfigure}[b]{0.5\linewidth}
    \centering
    \captionsetup{justification=centering}
    \includegraphics[width=1\linewidth]{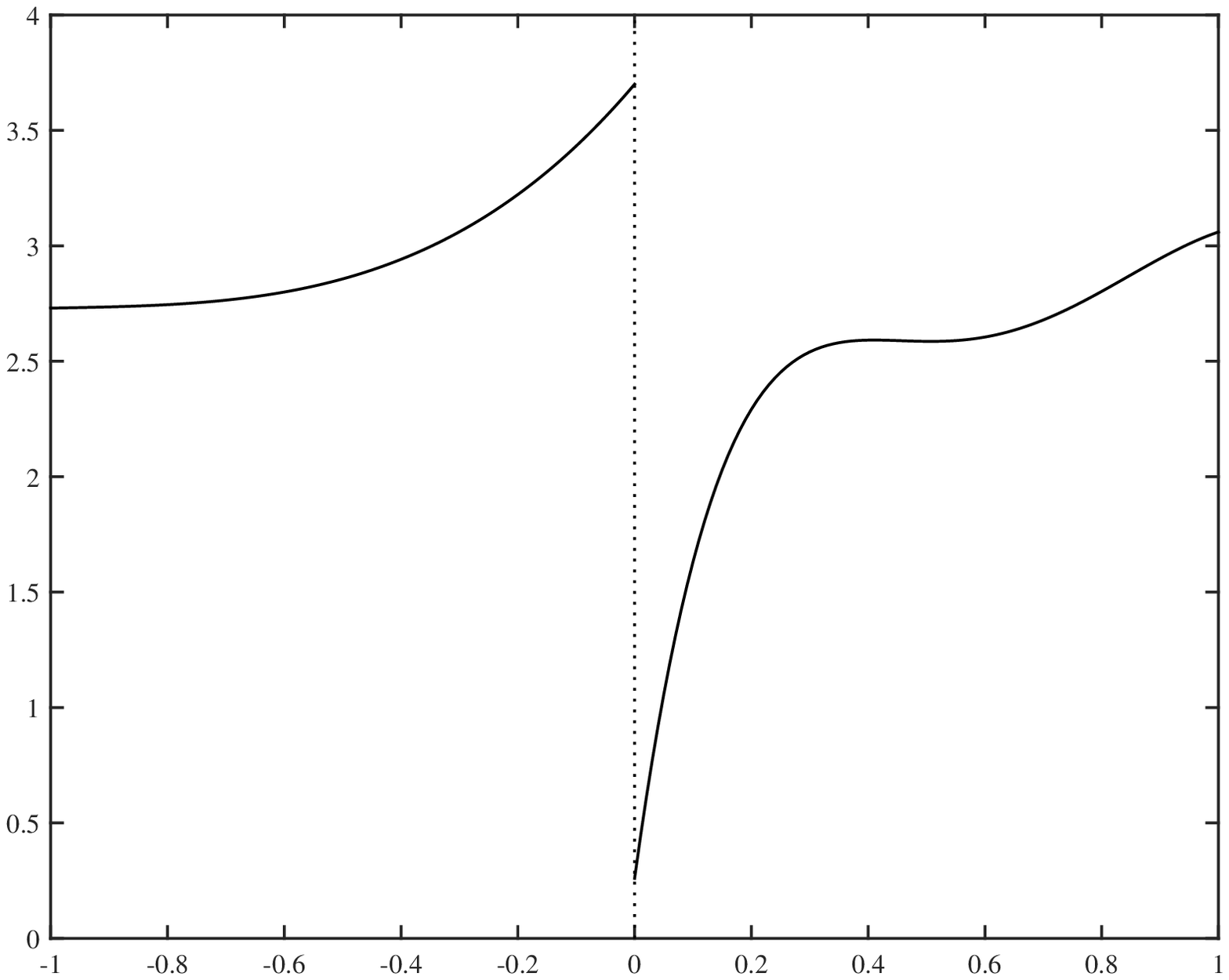} 
    \captionsetup{singlelinecheck=off}
    \caption[]
   {Design 2. Ludwig and Miller I (2007) Data (Design 2 of CCT)
   {\scriptsize
   \begin{align*}
    m_1(x) &= 0.26 + 18.49x - 54.8x^2 + 74.3x^3 - 45.02x^4 + 9.83x^5\\
    m_{0}(x) &= 3.70 + 2.99x + 3.28x^2 + 1.45x^3 + 0.22x^4 + 0.03x^5
    \end{align*}}}
    \label{figure:dgp:b} 
    \vspace{4ex}
  \end{subfigure} 
  \begin{subfigure}[b]{0.5\linewidth}
    \centering
    \captionsetup{justification=centering}
    \includegraphics[width=1\linewidth]{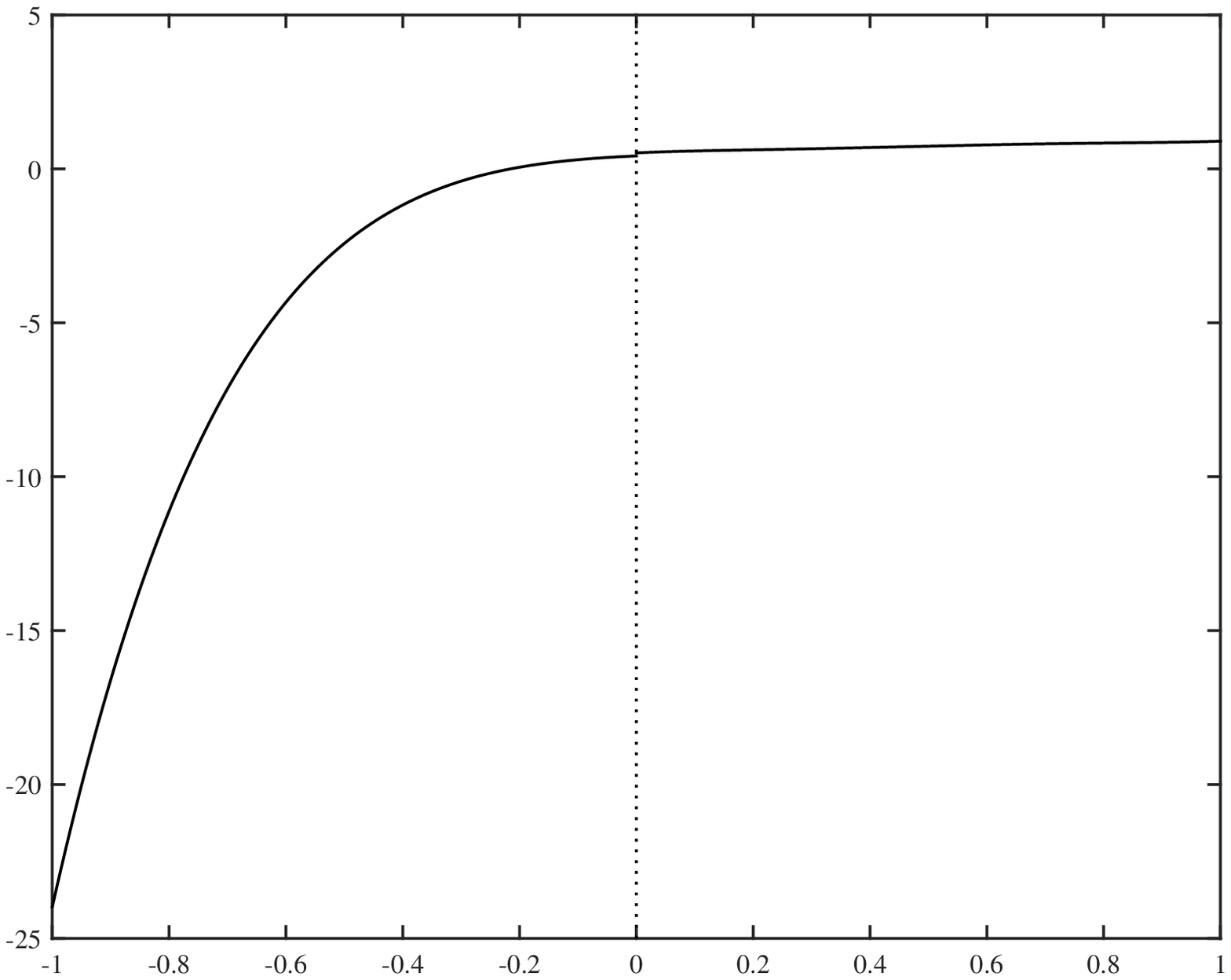} 
    \captionsetup{singlelinecheck=off}
    \caption[]
  {Design 3. Constant Additive Treatment Effect (Design 3 of IK)
   {\scriptsize
   \begin{align*}
    m_1(x) &= 0.52 + 0.84x - 3.0x^2 + 7.99x^3 - 9.01x^4 + 3.56x^5\\
    m_{0}(x) &= 0.42 + 0.84x - 3.0x^2 + 7.99x^3 - 9.01x^4 + 3.56x^5
   \end{align*}}}
    \label{figure:dgp:c} 
    \vspace{4ex}
  \end{subfigure}
  \hspace{5mm}
  \begin{subfigure}[b]{0.5\linewidth}
    \centering
    \captionsetup{justification=centering}
    \includegraphics[width=1\linewidth]{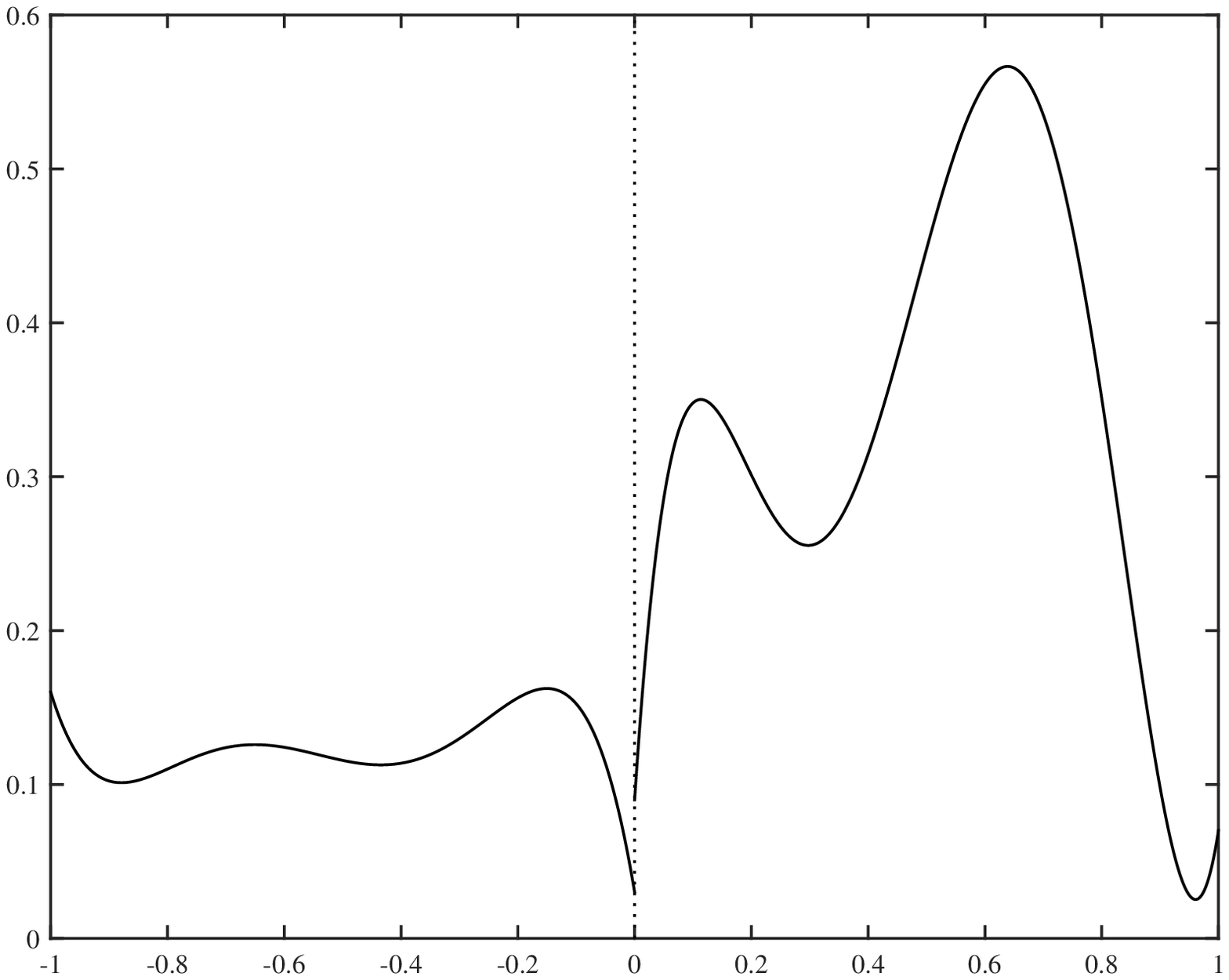} 
    \captionsetup{singlelinecheck=off}
    \caption[]
  {Design 4. Ludwig and Miller II (2007, Figure II. B) Data
   \hspace*{4mm}
   {\scriptsize 
   \begin{align*}
    m_1(x) &= 0.09 + 5.76 x - 42.56 x^2 + 120.90 x^3 - 139.71x^4 + 55.59 x^5\\
    m_{0}(z) &= 0.03 - 2.26 x - 13.14 x^2 -30.89 x^3  -31.98 x^4 -12.1 x^5
   \end{align*}}}
    \label{figure:dgp:d} 
    \vspace{4ex}
  \end{subfigure} 
  \caption{Simulation Design. The dashed line in the panel for Design 1 denotes the density of
the forcing variable. The supports for $m_{1}(x)$ and $m_{0}(x)$ are  $x\geq 0$ and $x<0$, respectively.}
  \label{figure:dgp} 
\end{figure}

The designs are depicted in Figure \ref{figure:dgp}. 
For the first two designs, the sign of the product of the second derivatives is negative so that the AMSE convergence rates for all bandwidth selection rules are the same. 
For the next two designs, the sign is positive.  
For these two cases, the AFO bandwidth has the faster convergence rate compared to IND.  Design 3, examined by IK, however, has the same second derivatives on the right and on the left of the cut of point, so that the convergence rate of the AMSE for IK is the same with that for the AFO.

For each design, the assignment variable $X_{i}$ is given by $2Z_{i}-1$ where $Z_{i}$ have a Beta distribution with parameters $\alpha=2$ and $\beta=4$. 
We consider a normally distributed additive error term with mean zero and standard deviation $0.1295$. 
The specification for the assignment variable and the additive error are exactly the same as that considered by IK. 
We use data sets of 500, 2,000 and 5,000 observations and the results are drawn from 10,000 replications.

\subsection{Results}

The simulation results are presented in Tables \ref{table:design1to2} and \ref{table:design3to4}. Table \ref{table:design1to2} reports the results for Designs 1 and 2. 
The first column explains the design and the second column shows the sample size. 
The third column reports the method used to obtain the bandwidth(s). MMSE refers to the feasible AFO bandwidth selection rule based on $MMSE_{n}^{p}(h)$ in equation (\ref{eq:LLRMMSEhatB}).
IND is the independent bandwidths. 
IK is the bandwidth denoted by $\hat{h}_{opt}$ in Table 2 of IK.\footnote{Algorithms provided by \citet{ik09} and IK differ slightly for computing
the variances and the regularization terms. See Section 4.2 of \citet{ik09} and Section 4.2 of IK for more details. 
Given that they provide a Stata code for the former and that it is used in many empirical researches, we show the result for the former. 
Our unreported simulation finds that two algorithms perform very similarly except Design 2 where the former
performs significantly better than the latter.} 
CV is the cross-validation bandwidth considered by \citet{lm05}; its implementation is described in Section 4.5 of IK. 
Note that the cross-validation bandwidth involves one ad hoc parameter which defines the neighborhood to compute the cross-validation criterion although other methods presented here are fully data-driven.\footnote{See Section 4.5 of IK for the ad hoc parameter $\delta$ used in the
cross-validation method to control the number of observation used to compute the criterion function. $\delta$ is set to 0.5 as in IK.} 
For the sample size of 2,000 and 5,000, we show results only for MMSE, IND, and IK due to computational burden.

The fourth and fifth columns report the mean (labeled `Mean') and standard deviation (labeled `SD') of the bandwidths for MMSE, IND, IK, and CV. 
For MMSE and IND, these columns report the bandwidth obtained for the right side of the cut-off point. 
The sixth and seventh columns report the corresponding ones on the left sides for MMSE and IND.
The eighth and ninth columns report the bias (Bias) and the root mean squared error (RMSE) for the sharp RDD estimate. 
Bias and RMSE are 5\% trimmed versions since unconditional finite sample variance of local linear estimators is infinite (see \citealp{sg96}). 
The tenth column report efficiency relative to the most efficient bandwidth selection rule based on the RMSE. 
The eleventh and twelfth columns report RMSE and efficiency based on the true objective functions for the respective bandwidth selection rules.
These can be considered as the theoretical predictions based on
asymptotic analysis.\footnote{A detailed procedure to obtain RMSE* is provided in the Supplemental Material.}

The sign of the product of the second derivatives is negative for Designs 1 and 2. 
Thus the AMSEs for all the bandwidth selection rules converge in the same rate, $n^{-4/5}$, where $n$ is the sample size.
The top panel of Table \ref{table:design1to2} reports the results for Design 1. 
For Design 1, theoretical efficiency is not so different across different bandwidth selection rules. 
Reflecting this, the simulation results show similar performances, for sample size 500, across different bandwidth selection rules. 
As the sample size increases, however, the performance of MMSE, which has the theoretical advantage in relative efficiency, dominates other methods. 
Note that the relative efficiency of the MMSE is higher than the asymptotic prediction for the sample sizes 2000 and 5000. 
This is attained by the finite sample performance of MMSE, in terms of
RMSE, realizing close to the theoretical prediction in these sample sizes, whereas other methods do not.

The bottom panel of Table \ref{table:design1to2} reports the results for Design 2. 
The magnitude of the ratio of the second derivatives is larger for this design compared with Design 1, so that the RMSE is larger for the same sample size. 
For Design 2, the results are very similar to the results for Design 1. 
Relative performance of IK is worse for this design compared to the performance in Design 1 reflecting the theoretical relative efficiency loss of IK for this design.

\begin{table}
\caption{Bias and RMSE for Designs 1 and 2}
\label{table:design1to2}
\begin{center}
\small
\begin{tabular}{cllccccccccc}
\hline \hline
&& & \multicolumn{2}{c}{$\hat h_{1}$} & \multicolumn{2}{c}{$\hat h_{0}$}  & \multicolumn{3}{c}{$\hat \tau$}\\ 
Design & $n$ & Method & Mean & SD & Mean & SD & Bias & RMSE & Eff & RMSE* & Eff*\\ \hline
1 & 500 & MMSE & 0.330 & 0.158 & 0.375 & 0.163 & 0.028 & 0.050 & 0.934 & 0.062 & 1 \\
 &  & IND & 0.765 & 0.572 & 0.588 & 0.402 & 0.039 & 0.047 & 1 & 0.063 & 0.981 \\
 &  & IK & 0.478 & 0.058 &  &  & 0.038 & 0.049 & 0.965 & 0.063 & 0.986 \\
 &  & CV & 0.416 & 0.093 &  &  & 0.037 & 0.048 & 0.975 \\ \cline{2-12}
 & 2000 & MMSE & 0.320 & 0.181 & 0.272 & 0.127 & 0.023 & 0.033 & 1 & 0.035 & 1 \\
 &  & IND & 0.730 & 0.604 & 0.359 & 0.120 & 0.041 & 0.042 & 0.837 & 0.036 & 0.979 \\
 &  & IK & 0.373 & 0.040 &  &  & 0.036 & 0.039 & 0.838 & 0.036 & 0.987 \\ \cline{2-12}
 & 5000 & MMSE & 0.280 & 0.177 & 0.181 & 0.081 & 0.018 & 0.025 & 1 & 0.025 & 1 \\
 &  & IND & 0.658 & 0.500 & 0.335 & 0.096 & 0.040 & 0.040 & 0.724 & 0.025 & 0.979 \\
 &  & IK & 0.339 & 0.034 &  &  & 0.032 & 0.034 & 0.723 & 0.025 & 0.987 \\ \hline
2 & 500 & MMSE & 0.075 & 0.005 & 0.188 & 0.041 & 0.039 & 0.074 & 1 & 0.081 & 1 \\
 &  & IND & 0.144 & 0.012 & 0.278 & 0.019 & 0.114 & 0.120 & 0.616 & 0.083 & 0.979 \\
 &  & IK & 0.249 & 0.016 &  &  & 0.138 & 0.142 & 0.521 & 0.088 & 0.913 \\
 &  & CV & 0.129 & 0.013 &  &  & 0.079 & 0.097 & 0.766 \\ \cline{2-12}
 & 2000 & MMSE & 0.055 & 0.002 & 0.138 & 0.010 & 0.021 & 0.041 & 1 & 0.046 & 1 \\
 &  & IND & 0.109 & 0.004 & 0.200 & 0.010 & 0.066 & 0.069 & 0.598 & 0.047 & 0.979 \\
 &  & IK & 0.178 & 0.005 &  &  & 0.069 & 0.072 & 0.568 & 0.051 & 0.913 \\ \cline{2-12}
 & 5000 & MMSE & 0.046 & 0.001 & 0.114 & 0.005 & 0.015 & 0.028 & 1 & 0.032 & 1 \\
 &  & IND & 0.086 & 0.002 & 0.163 & 0.007 & 0.043 & 0.045 & 0.628 & 0.033 & 0.979 \\
 &  & IK & 0.135 & 0.003 &  &  & 0.044 & 0.046 & 0.608 & 0.035 & 0.913 \\
 \hline
\end{tabular}
\caption*{\footnotesize Notes: $n$ is the sample size. ``Eff'' stands for the efficiency based on RMSE relative to MMSE. RMSE* and Eff* are based on the true objective functions for the respective bandwidth selection rules.}
\end{center}
\end{table}

Next, we turn to Designs 3 and 4, in which the sign of the product of the second derivatives is positive. 
In general, these cases should show the advantage of MMSE over IND, as the AMSE for it converges with rate $n^{-6/7}$ whereas IND's AMSE converges with rate $n^{-4/5}$.  
For Design 4 the same rate advantage holds for MMSE over IK.  
For Design 3, the second derivatives are the same and hence this is an exceptional case as discussed in section 2.2.

\begin{table}[htbp]
\caption{Bias and RMSE for Designs 3 and 4}
\label{table:design3to4}
\begin{center}
\small
\begin{tabular}{cllccccccccc}
\hline \hline
& & & \multicolumn{2}{c}{$\hat h_{1}$} & \multicolumn{2}{c}{$\hat h_{0}$}  & \multicolumn{3}{c}{$\hat \tau$}\\ 
Design& $n$ & Method & Mean & SD & Mean & SD & Bias & RMSE & Eff & RMSE* & Eff*\\ \hline
3 & 500 & MMSE & 0.312 & 0.160 & 0.207 & 0.052 & -0.023 & 0.052 & 0.954 & 0.046 & 1 \\
 &  & IND & 0.353 & 0.284 & 0.180 & 0.062 & -0.007 & 0.050 & 1 & 0.047 & 0.988 \\
 &  & IK & 0.174 & 0.016 &  &  & -0.014 & 0.050 & 0.986 & 0.046 & 0.998 \\
 &  & LM & 0.112 & 0.008 &  &  & -0.003 & 0.061 & 0.812 \\ \cline{2-12}
 & 2000 & MMSE & 0.303 & 0.157 & 0.167 & 0.032 & -0.009 & 0.026 & 1 & 0.026 & 1 \\
 &  & IND & 0.287 & 0.217 & 0.148 & 0.067 & -0.003 & 0.028 & 0.932 & 0.027 & 0.950 \\
 &  & IK & 0.140 & 0.013 &  &  & -0.007 & 0.028 & 0.935 & 0.026 & 0.999 \\ \cline{2-12}
 & 5000 & MMSE & 0.281 & 0.148 & 0.146 & 0.022 & -0.004 & 0.017 & 1 & 0.017 & 1 \\
 &  & IND & 0.250 & 0.225 & 0.127 & 0.059 & -0.001 & 0.019 & 0.889 & 0.019 & 0.925 \\
 &  & IK & 0.122 & 0.011 &  &  & -0.004 & 0.019 & 0.909 & 0.017 & 1 \\ \hline
4 & 500 & MMSE & 0.232 & 0.093 & 0.638 & 0.209 & -0.001 & 0.055 & 1 & 0.039 & 1 \\
 &  & IND & 0.605 & 0.530 & 1.210 & 0.965 & 0.058 & 0.062 & 0.888 & 0.072 & 0.545 \\
 &  & IK & 0.547 & 0.147 &  &  & 0.074 & 0.080 & 0.695 & 0.077 & 0.506 \\
 &  & LM & 0.306 & 0.195 &  &  & 0.055 & 0.070 & 0.791 \\ \cline{2-12}
 & 2000 & MMSE & 0.232 & 0.082 & 0.552 & 0.189 & 0.006 & 0.033 & 1 & 0.022 & 1 \\
 &  & IND & 0.525 & 0.438 & 0.990 & 0.835 & 0.054 & 0.055 & 0.609 & 0.041 & 0.530 \\
 &  & IK & 0.460 & 0.104 &  &  & 0.066 & 0.068 & 0.492 & 0.044 & 0.494 \\ \cline{2-12}
 & 5000 & MMSE & 0.224 & 0.081 & 0.495 & 0.183 & 0.008 & 0.024 & 1 & 0.015 & 1 \\
 &  & IND & 0.371 & 0.235 & 0.622 & 0.459 & 0.049 & 0.050 & 0.486 & 0.029 & 0.520 \\
 &  & IK & 0.351 & 0.078 &  &  & 0.054 & 0.055 & 0.435 & 0.031 & 0.485 \\
\hline
\end{tabular}
\caption*{\footnotesize Notes: $n$ is the sample size. ``Eff'' stands for the efficiency based on RMSE relative to MMSE. RMSE* and Eff* are based on the true objective functions for the respective bandwidth selection rules.}
\end{center}
\end{table}

The top panel of Table \ref{table:design3to4} shows the result for Design 3.
In this case, the IND and IK bandwidths bias terms cancel exactly and the AMSE for IK method indeed converges with rate $n^{-6/7}$ as discussed earlier. 
Under this design, while there are some variations for sample size 500,
the performances of all bandwidth selection rules match the asymptotic
theoretical predictions for sample sizes 2000 and 5000.

The bottom panel of Table \ref{table:design3to4} is the design in which the theoretical prediction of the performance of the MMSE clearly dominates other bandwidth selection rules. 
And the simulation results demonstrate this. 
IND, IK, and CV bandwidths tend to lead to larger biases, especially for sample sizes 2000 and 5000.

We emphasize here that the main advantage of using the feasible AFO bandwidth selection rule is to take advantage of situations like Design 4 without incurring much penalty in other cases. 
As we demonstrate by the simulations, in all cases the feasible AFO bandwidth selection rule's efficiency loss is not more than 7\%, while the gain in efficiency is more than 50\% for all other bandwidth selection rules.

\begin{figure}[ht] 
  \begin{subfigure}[b]{0.5\linewidth}
    \centering
    \includegraphics[width=1\linewidth]{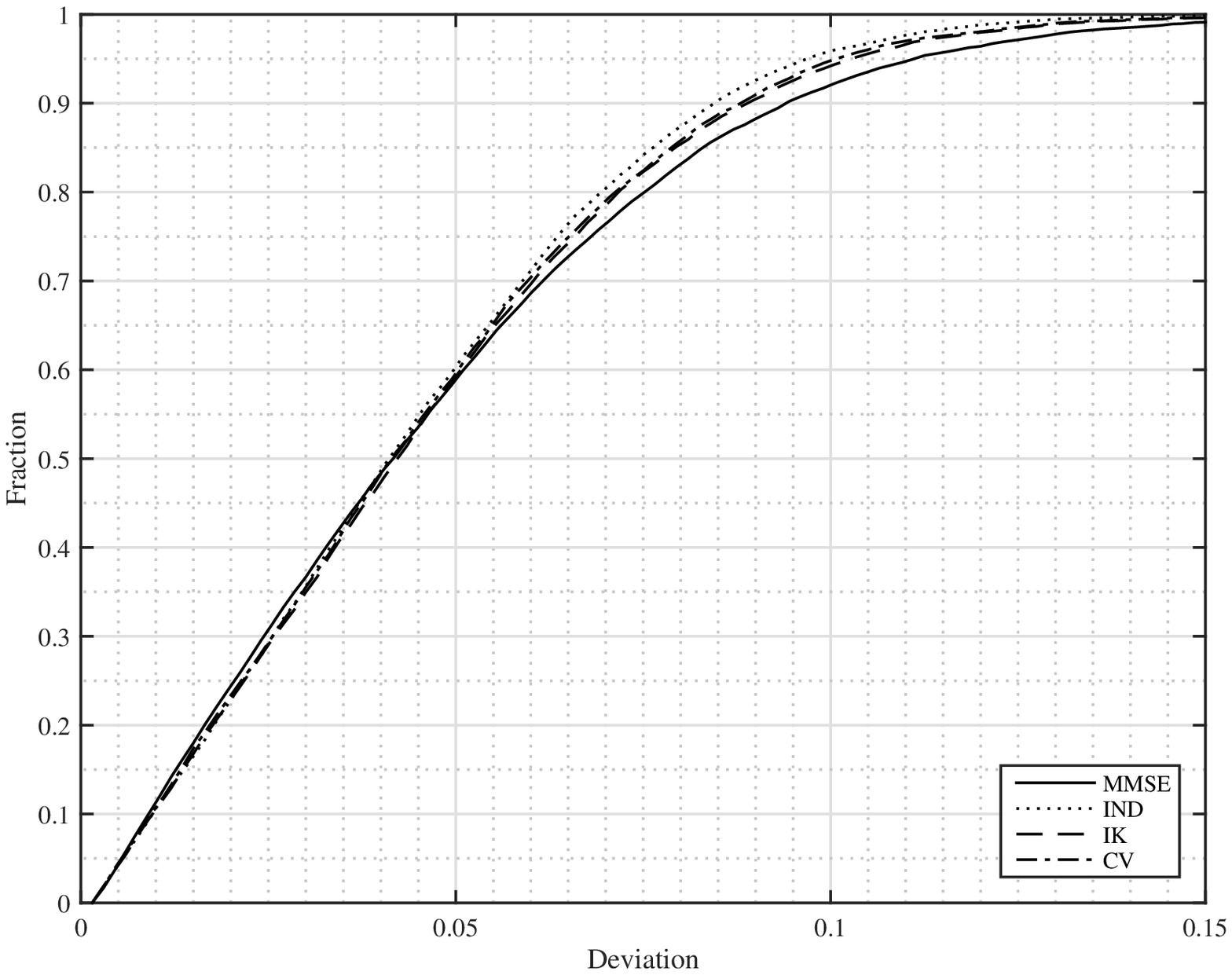} 
    \caption{Design 1, n=500} 
    \label{figure:coverage-base:12:1} 
    \vspace{4ex}
  \end{subfigure}
  \begin{subfigure}[b]{0.5\linewidth}
    \centering
    \includegraphics[width=1\linewidth]{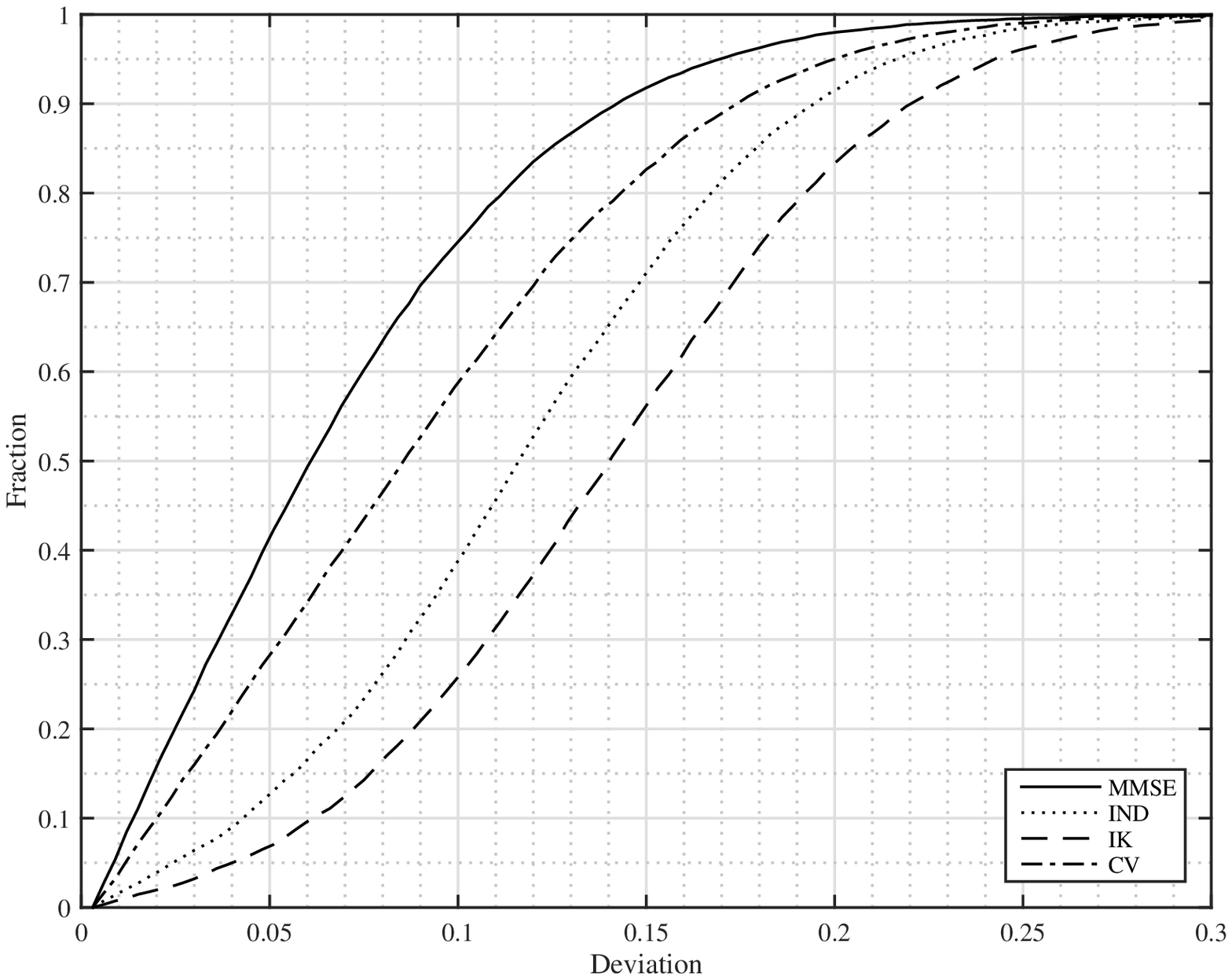} 
    \caption{Design 2, n=500} 
    \label{figure:coverage-base:12:2} 
    \vspace{4ex}
  \end{subfigure} 
  \begin{subfigure}[b]{0.5\linewidth}
    \centering
    \includegraphics[width=1\linewidth]{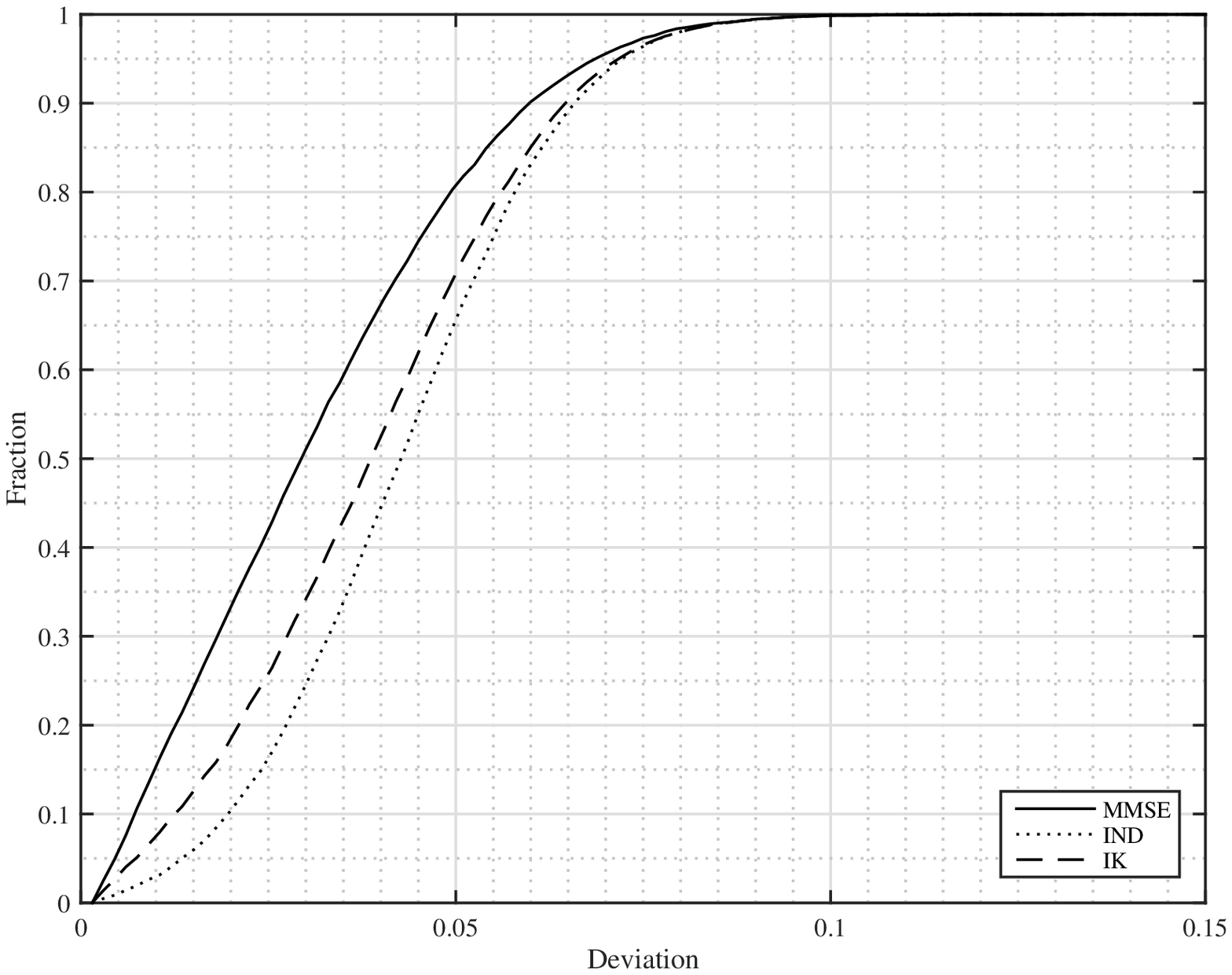} 
    \caption{Design 1, n=2,000} 
    \label{figure:coverage-base:12:1n2} 
    \vspace{4ex}
  \end{subfigure}
  \begin{subfigure}[b]{0.5\linewidth}
    \centering
    \includegraphics[width=1\linewidth]{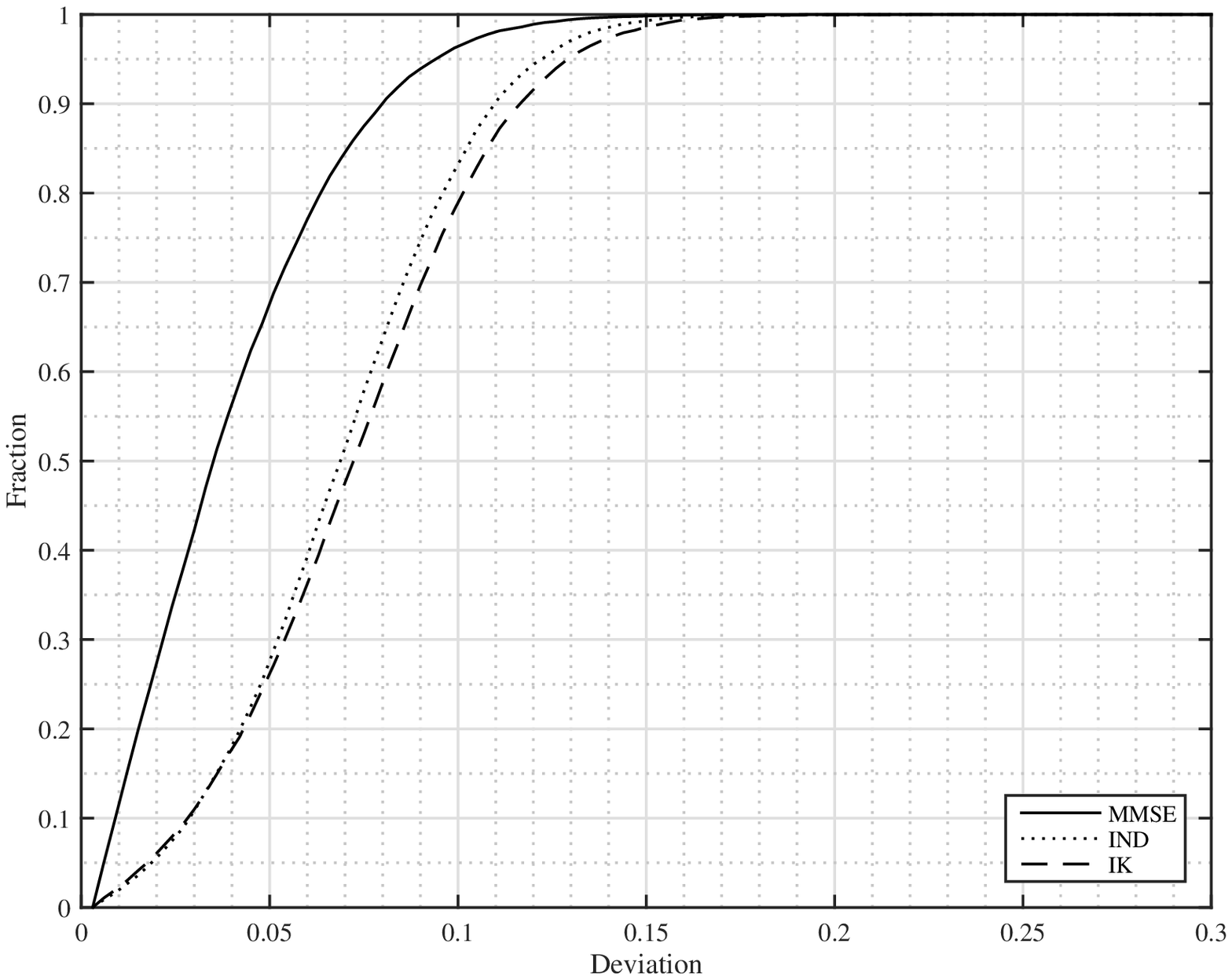} 
    \caption{Design 2, n=2,000} 
    \label{figure:coverage-base:12:2n2} 
    \vspace{4ex}
  \end{subfigure} 
    \begin{subfigure}[b]{0.5\linewidth}
    \centering
    \includegraphics[width=1\linewidth]{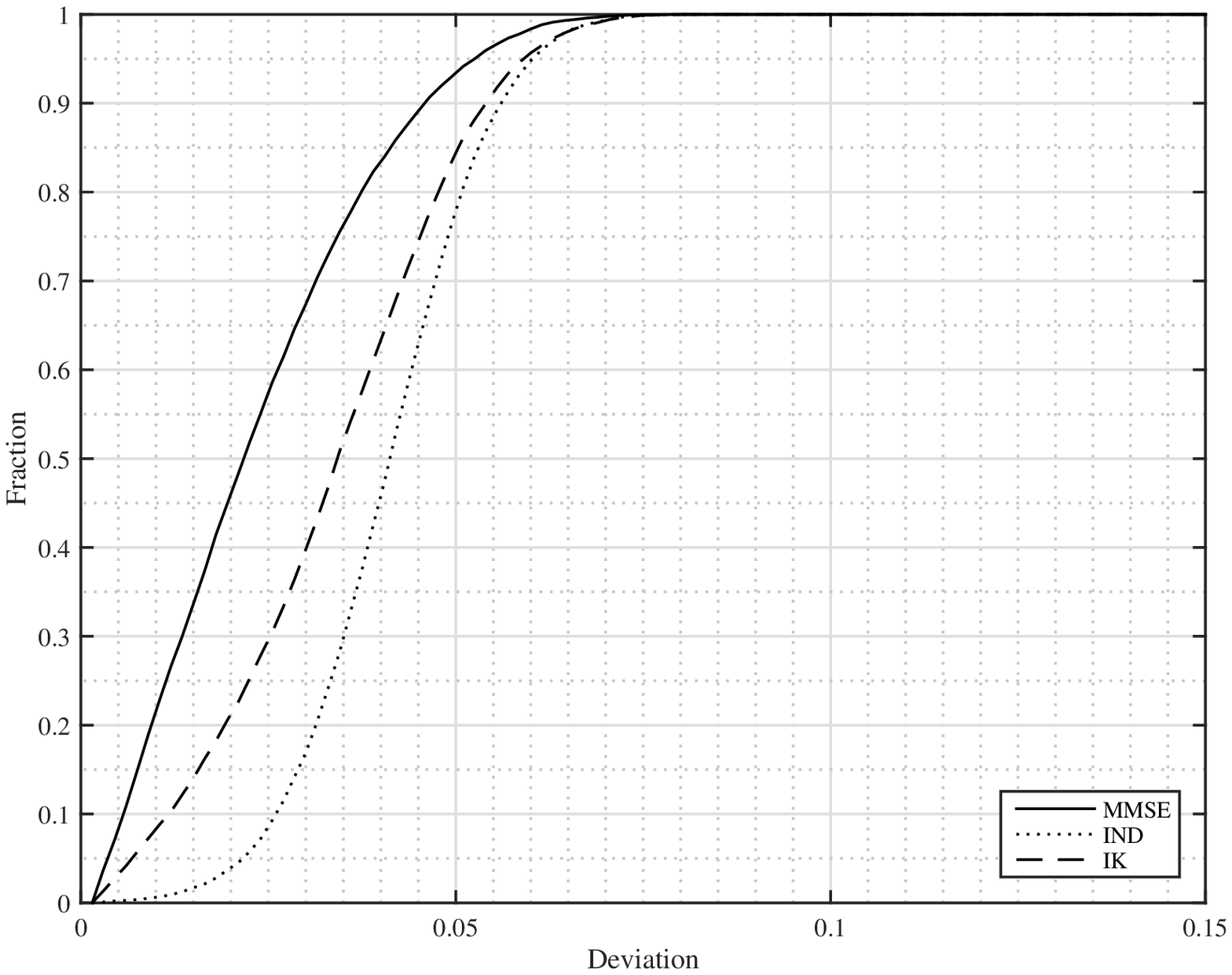} 
    \caption{Design 1, n=5,000} 
    \label{figure:coverage-base:12:1n5} 
    \vspace{4ex}
  \end{subfigure}
  \begin{subfigure}[b]{0.5\linewidth}
    \centering
    \includegraphics[width=1\linewidth]{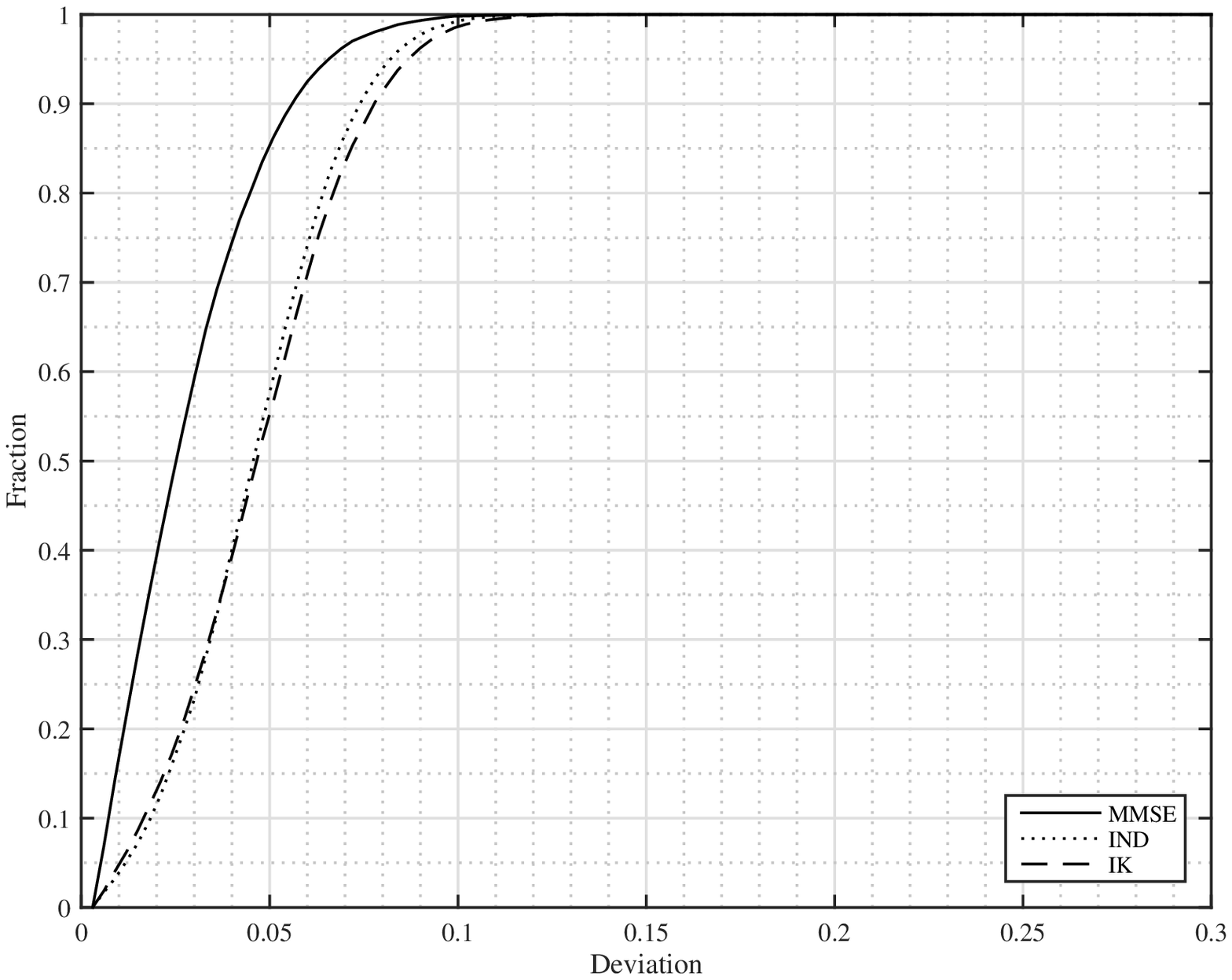} 
    \caption{Design 2, n=5,000} 
    \label{figure:coverage-base:12:2n5} 
    \vspace{4ex}
  \end{subfigure} 
\caption[]{Simulated CDF of $|\hat{\tau}-\tau|$ for different bandwidth selection rules for 10000 simulations}
  \label{figure:coverage-base:12} 
\end{figure}

\begin{figure}[ht] 
  \begin{subfigure}[b]{0.5\linewidth}
    \centering
    \includegraphics[width=1\linewidth]{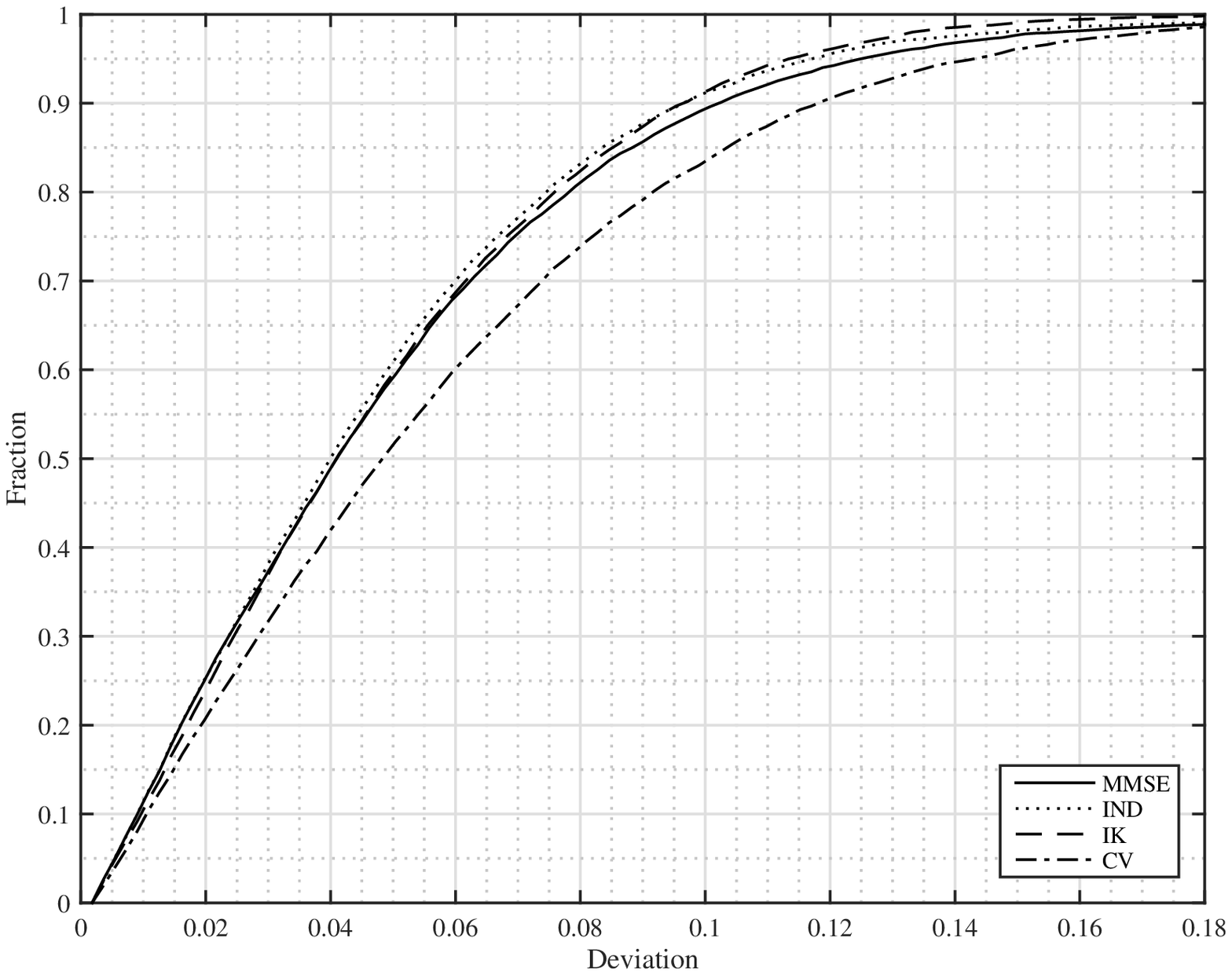} 
    \caption{Design 3, n=500} 
    \label{figure:coverage-base:34:3} 
    \vspace{4ex}
  \end{subfigure}
  \begin{subfigure}[b]{0.5\linewidth}
    \centering
    \includegraphics[width=1\linewidth]{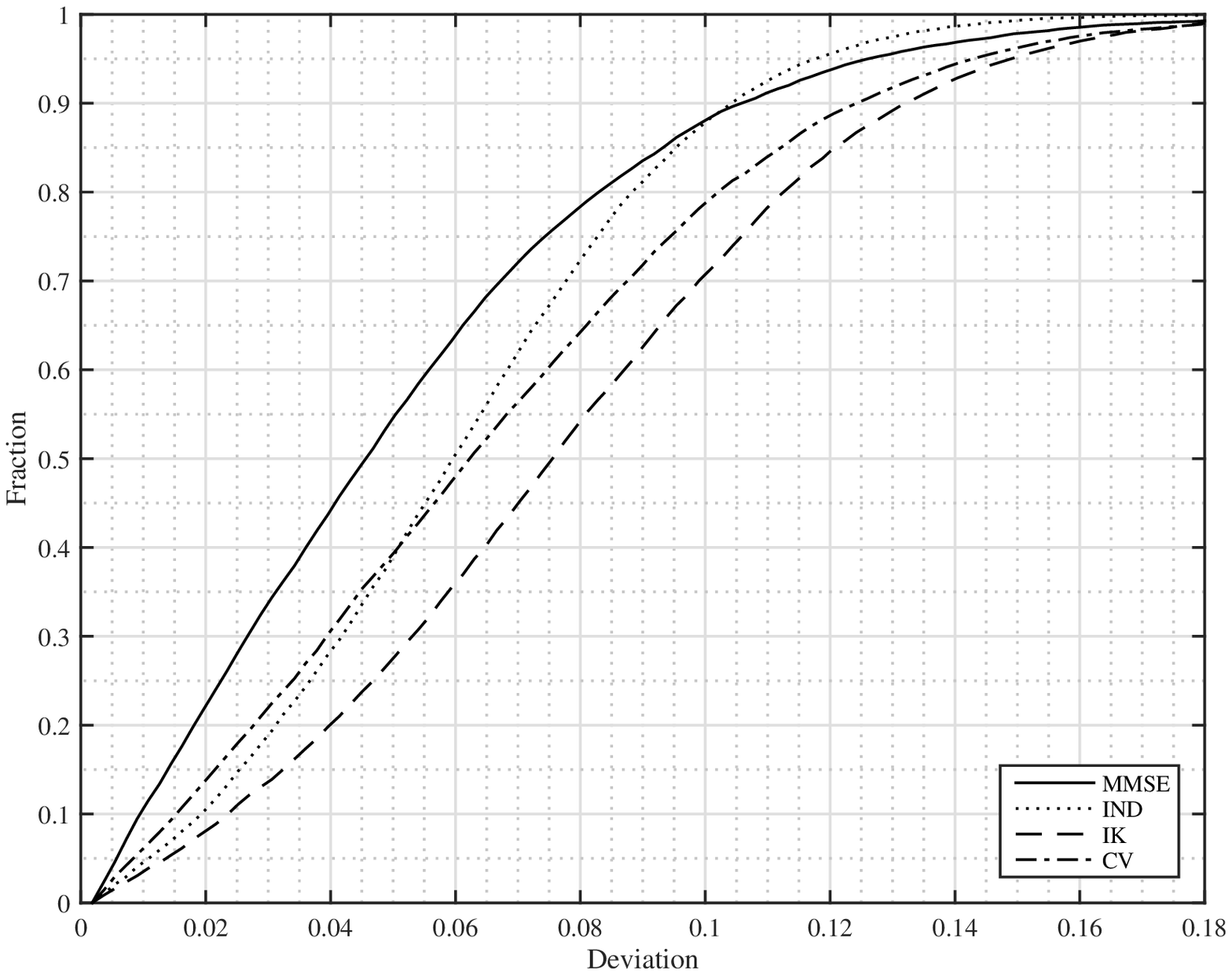} 
    \caption{Design 4, n=500} 
    \label{figure:coverage-base:34:4} 
    \vspace{4ex}
  \end{subfigure} 
  \begin{subfigure}[b]{0.5\linewidth}
    \centering
    \includegraphics[width=1\linewidth]{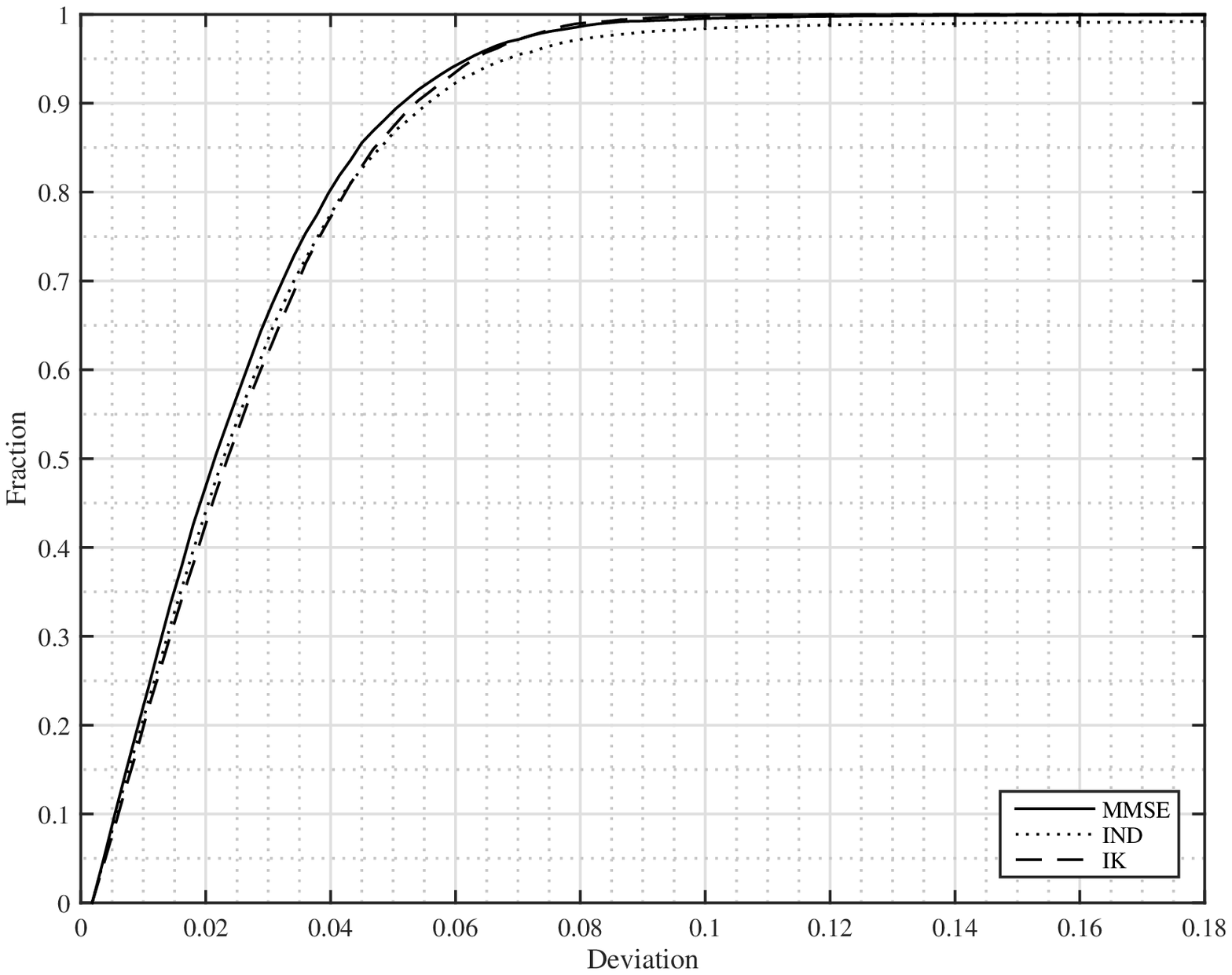} 
    \caption{Design 3, n=2,000} 
    \label{figure:coverage-base:34:3n2} 
    \vspace{4ex}
  \end{subfigure}
  \begin{subfigure}[b]{0.5\linewidth}
    \centering
    \includegraphics[width=1\linewidth]{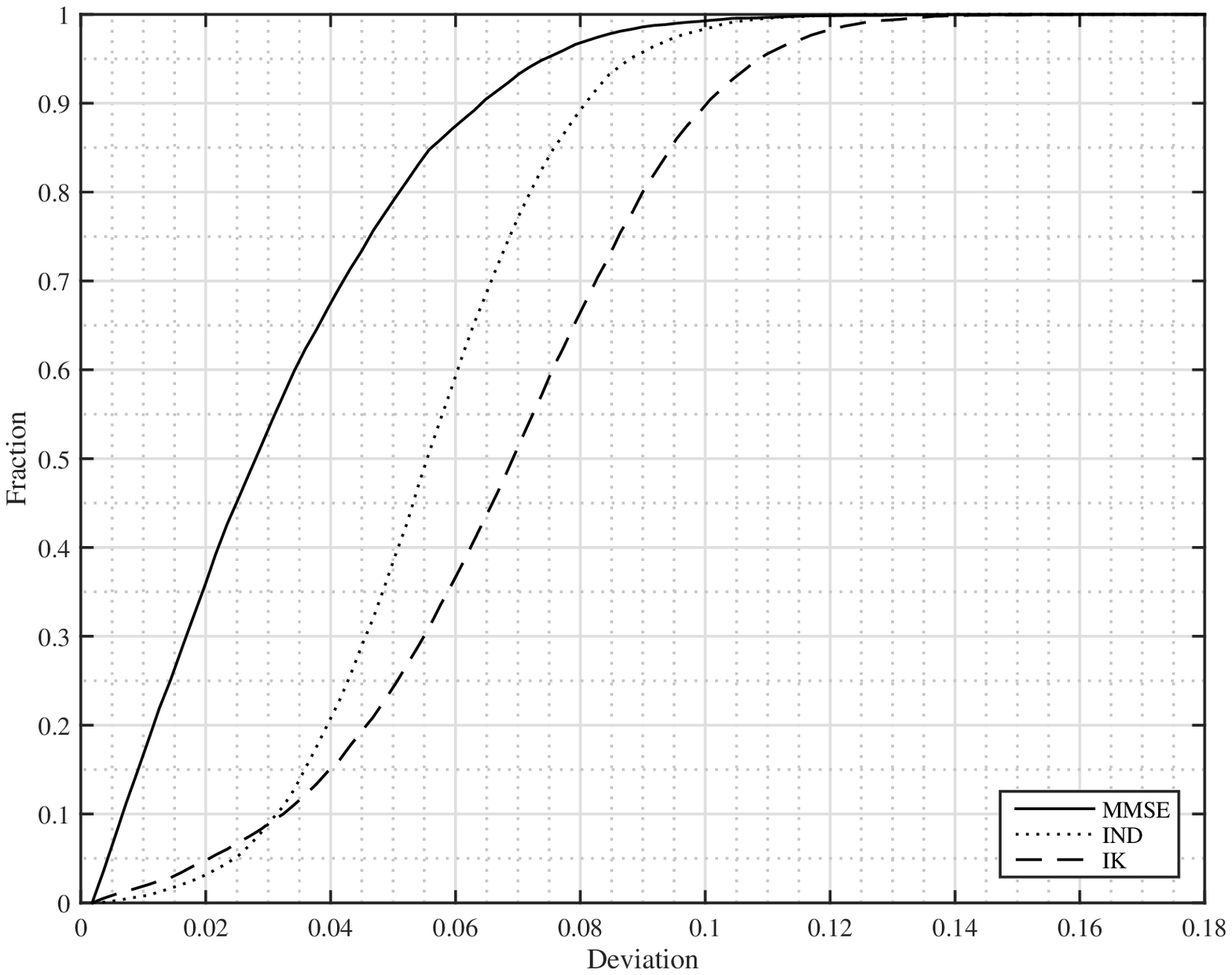} 
    \caption{Design 4, n=2,000} 
    \label{figure:coverage-base:34:4n2} 
    \vspace{4ex}
  \end{subfigure} 
    \begin{subfigure}[b]{0.5\linewidth}
    \centering
    \includegraphics[width=1\linewidth]{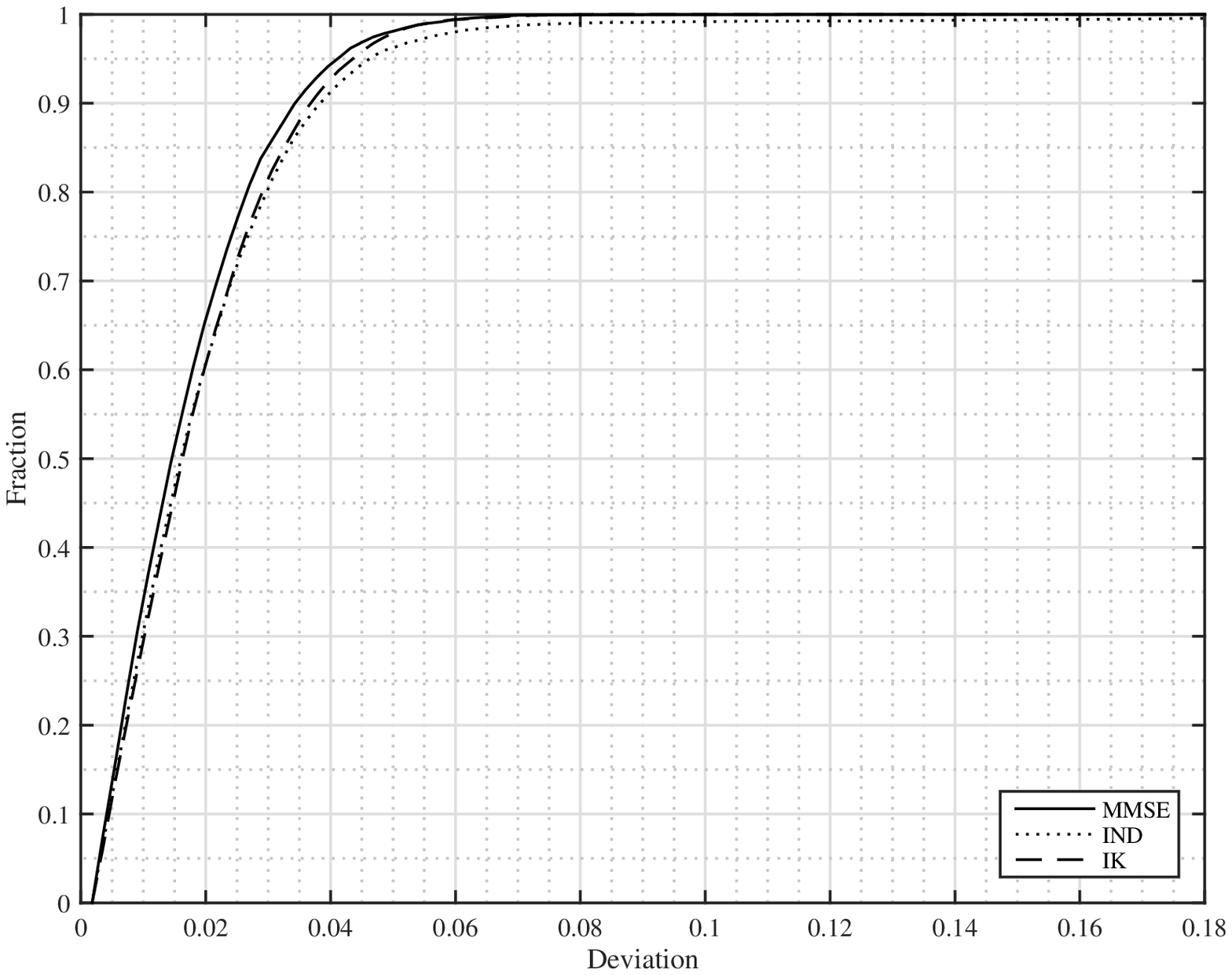} 
    \caption{Design 3, n=5,000} 
    \label{figure:coverage-base:34:3n5} 
    \vspace{4ex}
  \end{subfigure}
  \begin{subfigure}[b]{0.5\linewidth}
    \centering
    \includegraphics[width=1\linewidth]{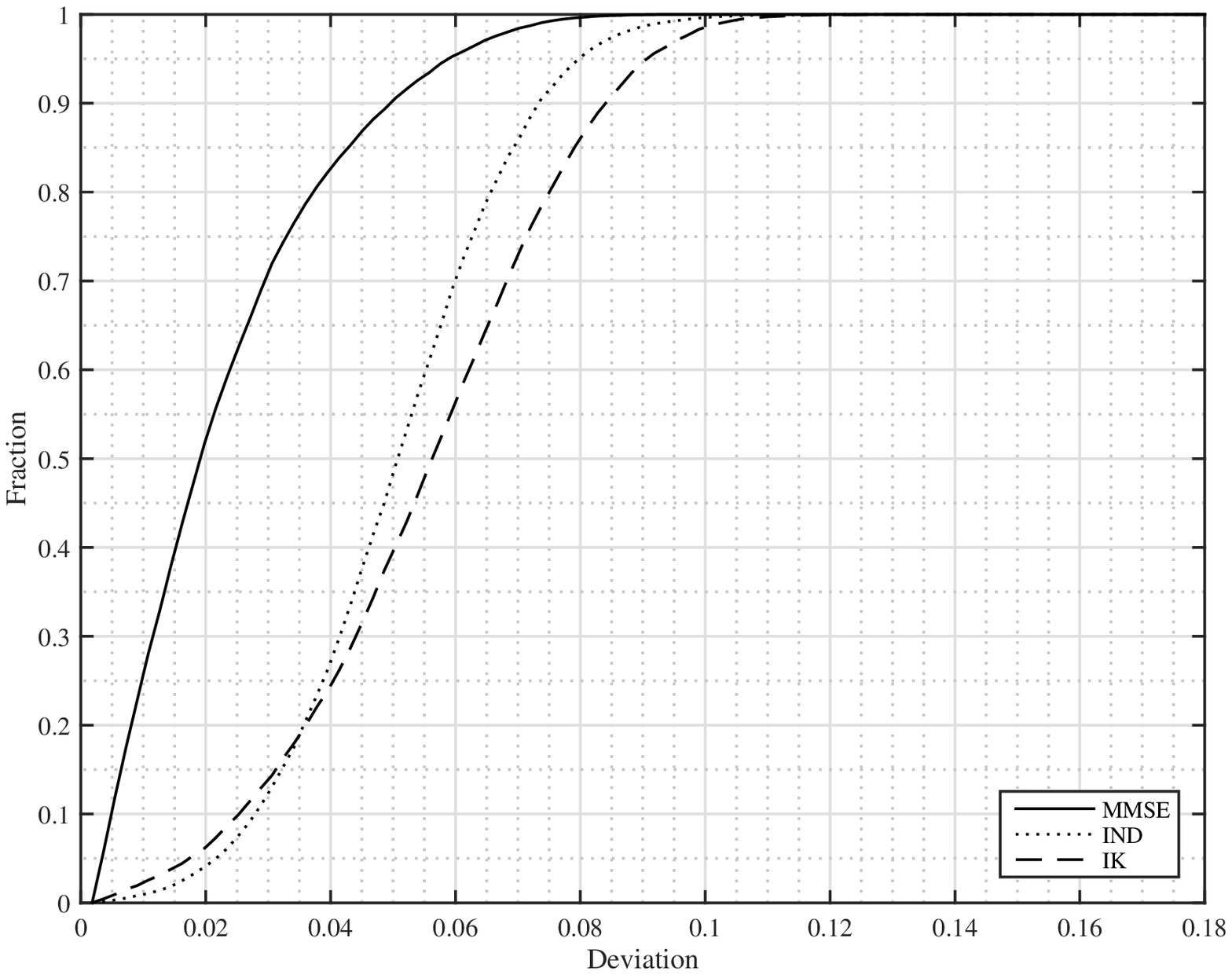} 
    \caption{Design 4, n=5,000} 
    \label{figure:coverage-base:34:4n5} 
    \vspace{4ex}
  \end{subfigure} 
\caption[]{Simulated CDF of $|\hat{\tau}-\tau|$ for different bandwidth selection rules for 10000 simulations}
  \label{figure:coverage-base:34} 
\end{figure}

We also demonstrate below that the comparison based on the RMSE can understate the difference between different bandwidth selection rules. 
This happens because large bias and very small variance can lead to reasonable size of the RMSE but this implies that RD estimators are concentrated on the biased value.
Figures \ref{figure:coverage-base:12} and \ref{figure:coverage-base:34} show the simulated CDF of $|\hat{\tau}-\tau|$ for different bandwidth selection rules for 10000 simulations. 

For Design 1, all methods work similarly with sample size 500. 
When sample size is increased to 2000 and 5000, the advantage of using the feasible AFO bandwidths over IND or IK bandwidths becomes evident. 
For example, the feasible AFO bandwidths dominates the IK bandwidth and the median deviation is
0.03 and 0.022 for the feasible AFO whereas it is 0.04 and 0.035 for
the IK bandwidth when the sample sizes are 2000 and 5000, respectively.

For Design 2, the advantage of the feasible AFO bandwidths is evident from the sample size 500.
For Design 3, there are very little differences across three methods as theory predicts.
For Design 4, the advantage of the feasible AFO bandwidths is again evident across all the sample sizes.

Finally, we show that the proposed method also estimates the conditional mean functions at the cut-off point reasonably well. 
The discussion provided in the previous section might have made an impression that the proposed method produce larger bias in estimating the conditional mean functions when the sign of the products of the second derivatives is positive while keeping the bias of the ``difference'' of the conditional mean functions small because removing the first-order bias term could incur larger bandwidths. 
This could be true if the second-order bias term doesn't work well as a penalty. 
Table \ref{table:m} reports the bias and the RMSE for the conditional mean functions, $m_{1}(c)$ and $m_{0}(c)$, at the cut-off point. 
There is no evidence that the proposed method estimates the RD parameter with larger bias of the estimates for the conditional mean functions.

\begin{table}[htbp]
\caption{Bias and RMSE for the Conditional Mean Functions, n=500}
\label{table:m}
\begin{center}
\small
\begin{tabular}{llcccc}
\hline \hline
& & \multicolumn{2}{c}{$\hat m_{1}(c)$} & \multicolumn{2}{c}{$\hat m_{0}(c)$}\\
Design & Method & Bias & RMSE & Bias & RMSE \\ \hline
Design 1 & MMSE & 0.007 & 0.037 & -0.021 & 0.037 \\
 & IK & 0.011 & 0.032 & -0.027 & 0.037 \\ \hline
Design 2 & MMSE & 0.028 & 0.071 & -0.010 & 0.040 \\
 & IK & 0.128 & 0.137 & -0.010 & 0.039 \\ \hline
Design 3 & MMSE & 0.005 & 0.038 & 0.029 & 0.056 \\
 & IK & 0.007 & 0.039 & 0.021 & 0.045 \\ \hline
Design 4 & MMSE & 0.094 & 0.110 & 0.095 & 0.098 \\
 & IK & 0.139 & 0.145 & 0.066 & 0.074 \\ \hline
\hline
\end{tabular}
\end{center}
\end{table}

In summary, the simulation results show that the feasible AFO bandwidth selection rule (MMSE) reproduces the theoretical predictions well so that when the AMSE is predicted to converge faster by theory compared to others, namely, when the sign of the product of the second derivatives is positive and the second derivatives are not the same, the feasible AFO bandwidth yield a better estimator of the treatment effect at the threshold as shown in Tables \ref{table:design1to2} and \ref{table:design3to4}, and Figures \ref{figure:coverage-base:12}  and \ref{figure:coverage-base:34}. 
Moreover, when the sign of the product of the second derivatives is negative, even in cases where the asymptotic gain is expected to be modest, the performance of the feasible AFO can be much better than others (see Design 2), because the feasible AFO bandwidth performs according to asymptotic theory, whereas other methods do not. 
Overall, MMSE appears very promising.
The proposed method performs significantly better than existing methods for cases that are not artificial but motivated by the empirical researches.

\section{Empirical Illustration}

We illustrate how the proposed method in this paper can contribute to empirical researches. In doing so, we revisit the problem considered by \citet{lm07}. 
They investigate the effect of Head Start (hereafter HS) on health and schooling. HS is the federal government's program aimed to provide preschool, health, and other social services to poor children age three to five and their families. 
They note that the federal government assisted HS proposals of the 300 poorest counties based on the county's 1960 poverty rate and find that the county's 1960 poverty rate can become the assignment variable where the cut-off value is given by 59.1984.\footnote{Since the poverty rate is based on the county level information, the sampling framework does not exactly correspond to the one considered in the paper. However, in this illustration we follow the estimation framework used by \citet{lm07} which fits into our framework.} 
They assess the effect of HS assistance on numerous measures such as HS participation, HS spending, other social spending, health, mortality and education.

Here we revisit the study on the effect of HS assistance on HS spending and mortality provided in Tables II and III of \citet{lm07}. 
The outcome variables considered in Tables II and III include HS spending per child in 1968 and 1972, and the mortality rate for the causes of death that could be affected by the Head Start health services to all and black children age five to nine. 
1972 HS spending per child and the mortality rate for all children generated the simulation Designs 2 and 4 in the previous section, respectively. 
In obtaining the RD estimates, they employ the LLR using a triangular kernel function as proposed by \citet{por03}. 
For bandwidths, they use 3 different bandwidths, 9, 18 and 36 in somewhat ad-hoc manner rather than relying on some bandwidths selection methods. 
This implies that the bandwidths and the number of observations with nonzero weight used for estimation are independent of outcome variables.

Columns 3 to 5 in Table \ref{table:mmseresults} reproduce the results presented in Tables II and III of \citet{lm07} for comparison.
The point estimates for 1968 HS spending per child range from 114.711 to
137.251. Perhaps we may say that they are not very sensitive to the
choice of bandwidth in this case. 
However, the point estimates for 1972 HS spending per child range from 88.959 to 182.396. 
What is more troubling would be the fact that they produce mixed results in statistical significance.
For 1968 HS spending per child, the point estimate with the bandwidth of
36 produce the result which is statistically significant at 5\% level
while the estimates with bandwidths of 9 and 18 are not statistically significant even at 10\% level. 
The results for 1972 HS spending per child are similar in the sense that
the estimates based on the
bandwidths of 9 and 36 are statistically significant at 10\% level while the estimate based on the bandwidth of 18 is not at the same level.

The results on the mortality rate for all children five to nine exhibit statistical significance though the point estimates range from -1.895 to -1.114 depending on which bandwidth to employ. 
The point estimate for the mortality rate for black children five to nine with bandwidth 18 is -2.719 which is statistically significant at 5\% level while the point estimates with bandwidths 9 and 36 are -2.275 and -1.589, respectively, which are not statistically insignificant even at 10\% level. 
It would be meaningful to see what sophisticated bandwidth selection methods can offer under situations where the results based on ad-hoc approaches cannot be interpreted easily.

Columns 6 and 7 in Table \ref{table:mmseresults} present the result based on the bandwidth selection methods based on MMSE and IK. 
For 1968 HS spending per child, the point estimates based on both
methods are similar but statistically insignificant although MMSE
produces a smaller standard error reflecting the larger bandwidth on the left of the cut-off. 
The point estimate for 1972 HS spending per child differ substantially although they are not statistically significant.
For the mortality rate for all children five to nine, both methods produce similar results in terms of the point estimates as well as statistical significance while they generate very different results in both point estimate and statistical significance for black children. 
To summarize, we found large but statistically insignificant point estimates for HS spending and statistically significant estimates for mortality rates by the proposed method in this paper. 
The results presented in Table \ref{table:mmseresults} alone do not imply any superiority of the proposed method over the existing methods because we never know true causal relationships. 
However, the results based on the proposed method should provide a meaningful perspective given the simulation experiments demonstrated in the previous section.

\begin{landscape}
\begin{table}
\caption{RD Estimates of the Effect of Head Start Assistance}
\label{table:mmseresults}
\begin{center}
\small
\begin{tabular}{llccccc}
\hline \hline
Variable & & & LM & & MMSE & IK \\ \hline
1968 HS spending & Bandwidth & 9 & 18 & 36 & $\langle 26.237, 45.925\rangle$ & 19.013 \\
 & Number of obs. & [217, 310] & [287, 674] & [300, 1877]  & [299, 2633] & [290, 727] \\
& RD estimate & 137.251 & 114.711 & $134.491^{**}$ & 110.590 & $108.128$ \\
& SE & (128.968) & (91.267) & (62.593) & (76.102) & (80.179) \\ \hline
1972 HS spending &  Bandwidth & 9 & 18 & 36 & $\langle 22.669, 42.943\rangle$ & 20.924 \\
 &  Number of obs. & [217, 310] & [287, 674] & [300, 1877] & [298, 2414] & [294, 824] \\
&  RD estimate & $182.119^{*}$ & 88.959 & $130.153^{*}$ & 105.832 & 89.102 \\
&  SE & (148.321) & (101.697) & (67.613) & (79.733) & (84.0272) \\ \hline
Child mortality, All & Bandwidth & 9 & 18 & 36 & $\langle 8.038, 14.113\rangle$ & 7.074 \\
& Number of obs. & [217, 310] & [287, 674] & [300, 1877] & [203, 508] & [184, 243] \\
& RD estimate & $-1.895^{**}$ & $-1.198^{*}$ & $-1.114^{**}$ & $-2.094^{***}$ & $-2.3589^{***}$ \\
& SE & (0.980) & (0.796) & (0.544) & (0.606) & (0.822) \\ \hline
Child mortality, Black & Bandwidth & 9 & 18 & 36 & $\langle 22.290, 25.924\rangle$ & 9.832 \\
& Number of obs. & [217, 310] & [287, 674] & [300, 1877] & [266, 968] & [209, 312] \\
& RD estimate & $-2.275$ & $-2.719^{**}$ & $-1.589$ & $-2.676^{***}$ & $-1.394$ \\
& & (3.758) & (2.163) & (1.706) & (1.164) & (2.191) \\ \hline
\hline
\end{tabular}
 \caption*{\footnotesize This result for LM is reproduced based on Tables II and III of \citet{lm07}. The bandwidths on the right and the left of the cut-off points are presented in angle brackets. The numbers of observations with nonzero weight on the right and the left of the cut-off are shown in square brackets. Standard errors are presented in parentheses.  ***, ** and * indicate statistical significance based on the bias-corrected $t$-value at 1\%, 5\% and 10\% level, respectively. See Equation (4.5), (4.7) and (4.8) in \citet[Section 4.3]{fg96} for estimation of the bias and standard error.}
\end{center}
\end{table}
\end{landscape}

\section{Conclusion}
\label{sec:discussion} 
In this paper, we have proposed a new bandwidth selection method for the RD estimators. 
We provided a discussion on the validity of the simultaneous choice of the bandwidths theoretically and illustrated that the proposed bandwidths produce results comparable to the theoretical results in the sample sizes relevant for empirical works.

A main feature of the proposed method is that we choose two bandwidths simultaneously. 
When we allow two bandwidths to be distinct, we showed that the minimization problem of the AMSE exhibits dichotomous characteristics depending on the sign of the product of the second derivatives of the underlying functions and that the optimal bandwidths that minimize the AMSE are not well-defined when the sign of the product is positive.
We introduced the concept of the AFO bandwidths and proposed a feasible version of the AFO bandwidths. The feasible bandwidths are proved to be asymptotically as good as the AFO bandwidths. 
A simulation study based on designs motivated by existing empirical literatures exhibits non-negligible gain of the proposed method under the situations where a single-bandwidth approach can become quite misleading. We also demonstrated how the proposed method can be implemented via an empirical example.

\citet{cct14} proposes robust confidence intervals for both sharp and fuzzy RD designs. The bandwidth selection rule we discuss in this paper may be used in a similar way to construct a robust confidence intervals. 
However, the extension is non-trivial as it is desirable to find a method that does not require knowing the sign of the product of the second derivatives of the regression function at both sides of the cut-off point. This is left as a future research.

\appendix

\section{Proofs of Theorem \ref{theorem:LLRpluginB}}

Recall that the objective function is: 
\begin{align*}
 & MMSE_{n}^{p}(h)=\left\{ \frac{b_{1}}{2}\left[\hat{m}_{1}^{(2)}(c)h_{1}^{2}-\hat{m}_{0}^{(2)}(c)h_{0}^{2}\right]\right\} ^{2}+\left[\hat{b}_{2,1}(c)h_{1}^{3}-\hat{b}_{2,0}(c)h_{0}^{3}\right]^{2}\\
 & \hspace{65mm}+\frac{\nu}{n\hat{f}(c)}\left\{ \frac{\hat{\sigma}_{1}^{2}(c)}{h_{1}}+\frac{\hat{\sigma}_{0}^{2}(c)}{h_{0}}\right\} .
\end{align*}

To begin with, we show that $\hat{h}_{1}$ and $\hat{h}_{0}$ satisfy
Assumption \ref{assumption:bandwidth}. If we choose a sequence of
$h_{1}$ and $h_{0}$ to satisfy Assumption \ref{assumption:bandwidth},
then $MMSE_{n}^{p}(h)$ converges to 0. Assume to the contrary that
either one or both of $\hat{h}_{1}$ and $\hat{h}_{0}$ do not satisfy
Assumption \ref{assumption:bandwidth}. Since $m_{0}^{(2)}(c)^{3}b_{2,1}(c)^{2}\neq m_{1}^{(2)}(c)^{3}b_{2,0}(c)^{2}$
by assumption, $\hat{m}_{0}^{(2)}(c)^{3}\hat{b}_{2,1}(c)^{2}\neq\hat{m}_{1}^{(2)}(c)^{3}\hat{b}_{2,0}(c)^{2}$
with probability approaching 1. Without loss of generality, we assume
this as well. Then at least one of the first-order bias term, the
second-order bias term and the variance term of $MMSE_{n}^{p}(\hat{h})$
does not converge to zero in probability. Then $MMSE_{n}^{p}(\hat{h})>MMSE_{n}^{p}(h)$
holds for some $n$. This contradicts the definition of $\hat{h}$.
Hence $\hat{h}$ satisfies Assumption \ref{assumption:bandwidth}.

We first consider the case in which $m_{1}^{(2)}(c)m_{0}^{(2)}(c)<0$.
In this case, with probability approaching 1, $\hat{m}_{1}^{(2)}(c)\hat{m}_{0}^{(2)}(c)<0$,
so that we assume this without loss of generality. When this holds,
note that the leading terms are the first term and the last term of
$MMSE_{n}^{p}(\hat{h})$ since $\hat{h}_{1}$ and $\hat{h}_{0}$ satisfy
Assumption \ref{assumption:bandwidth}. Define the plug-in version
of $AMSE_{1n}(h)$ provided in Definition \ref{def:FirstOptimalRegB}
by 
\[
AMSE_{1n}^{p}(h)=\left\{ \frac{b_{1}}{2}\left[\hat{m}_{1}^{(2)}(c)h_{1}^{2}-\hat{m}_{0}^{(2)}(c)h_{0}^{2}\right]\right\} ^{2}+\frac{\nu}{n\hat{f}(c)}\left\{ \frac{\hat{\sigma}_{1}^{2}(c)}{h_{1}}+\frac{\hat{\sigma}_{0}^{2}(c)}{h_{0}}\right\} .
\]
A calculation yields $\tilde{h}_{1}=\tilde{C}_{1}n^{-1/5}$ and $\tilde{h}_{0}=\tilde{C}_{0}n^{-1/5}$
where 
\begin{align*}
\tilde{C}_{1} & =\left\{ \frac{v\hat{\sigma}_{1}^{2}(c)}{b_{1}^{2}\hat{f}(c)\hat{m}_{1}^{(2)}(c)\left[\hat{m}_{1}^{(2)}(c)-{\hat{\lambda}_{1}}^{2}\hat{m}_{0}^{(2)}(c)\right]}\right\} ^{1/5},\quad\hat{\lambda}_{1}=\left\{ -\frac{\hat{\sigma}_{0}^{2}(c)\hat{m}_{1}^{(2)}(c)}{\hat{\sigma}_{1}^{2}(c)\hat{m}_{0}^{(2)}(c)}\right\} ^{1/3},
\end{align*}
and $\tilde{C}_{0}=\tilde{C}_{1}\hat{\lambda}_{1}$. With this choice,
$AMSE_{1n}^{p}(\tilde{h})$ and hence $MMSE_{n}^{p}(\tilde{h})$ converges
at the rate of $n^{-4/5}$. Note that if $\hat{h}_{1}$ or $\hat{h}_{0}$
converges at the rate slower than $n^{-1/5}$, then the bias term
converges at the rate slower than $n^{-4/5}$. If $\hat{h}_{1}$ or
$\hat{h}_{0}$ converges at the rate faster than $n^{-1/5}$, then
the variance term converges at the rate slower than $n^{-4/5}$. Thus
the minimizer of $MMSE_{n}^{p}(h)$, $\hat{h}_{1}$ and $\hat{h}_{0}$
converges to 0 at rate $n^{-1/5}$.

Thus we can write $\hat{h}_{1}=\hat{C}_{1}n^{-1/5}+o_{p}(n^{-1/5})$
and $\hat{h}_{0}=\hat{C}_{0}n^{-1/5}+o_{p}(n^{-1/5})$ for some $O_{P}(1)$
sequences $\hat{C}_{1}$ and $\hat{C}_{0}$ that are bounded away
from 0 and $\infty$ as $n\rightarrow\infty$. Using this expression,
\begin{eqnarray*}
MMSE_{n}^{p}(\hat{h}) & = & n^{-4/5}\left\{ \frac{b_{1}}{2}\left[\hat{m}_{1}^{(2)}(c)\hat{C}_{1}^{2}-\hat{m}_{0}^{(2)}(c)\hat{C}_{0}^{2}\right]\right\} ^{2}\\
 &  & +\frac{\nu}{n^{4/5}\hat{f}(c)}\left\{ \frac{\hat{\sigma}_{1}^{2}(c)}{\hat{C}_{1}}+\frac{\hat{\sigma}_{0}^{2}(c)}{\hat{C}_{0}}\right\} +o_{p}(n^{-4/5}).
\end{eqnarray*}
Note that 
\[
MMSE_{n}^{p}(\tilde{h})=n^{-4/5}\left\{ \frac{b_{1}}{2}\left[\hat{m}_{1}^{(2)}(c)\tilde{C}_{1}^{2}-\hat{m}_{0}^{(2)}(c)\tilde{C}_{0}^{2}\right]\right\} ^{2}+\frac{\nu}{n^{4/5}\hat{f}(c)}\left\{ \frac{\hat{\sigma}_{1}^{2}(c)}{\tilde{C}_{1}}+\frac{\hat{\sigma}_{0}^{2}(c)}{\tilde{C}_{0}}\right\} +O_{P}(n^{-8/5}).
\]
Since $\hat{h}$ is the optimizer, $MMSE_{n}^{p}(\hat{h})/MMSE_{n}^{p}(\tilde{h})\leq1$.
Thus 
\[
\frac{\left\{ \frac{b_{1}}{2}\left[\hat{m}_{1}^{(2)}(c)\hat{C}_{1}^{2}-\hat{m}_{0}^{(2)}(c)\hat{C}_{0}^{2}\right]\right\} ^{2}+\frac{\nu}{\hat{f}(c)}\left\{ \frac{\hat{\sigma}_{1}^{2}(c)}{\hat{C}_{1}}+\frac{\hat{\sigma}_{0}^{2}(c)}{\hat{C}_{0}}\right\} +o_{p}(1)}{\left\{ \frac{b_{1}}{2}\left[\hat{m}_{1}^{(2)}(c)\tilde{C}_{1}^{2}-\hat{m}_{0}^{(2)}(c)\tilde{C}_{0}^{2}\right]\right\} ^{2}+\frac{\nu}{\hat{f}(c)}\left\{ \frac{\hat{\sigma}_{1}^{2}(c)}{\tilde{C}_{1}}+\frac{\hat{\sigma}_{0}^{2}(c)}{\tilde{C}_{0}}\right\} +O_{P}(n^{-4/5})}\leq1.
\]

Note that the denominator converges to 
\[
\left\{ \frac{b_{1}}{2}\left[m_{1}^{(2)}(c)C_{1}^{*2}-m_{0}^{(2)}(c)C_{0}^{*2}\right]\right\} ^{2}+\frac{\nu}{f(c)}\left\{ \frac{\sigma_{1}^{2}(c)}{C_{1}^{*}}+\frac{\sigma_{0}^{2}(c)}{C_{0}^{*}}\right\} ,
\]
where $C_{1}^{*}$ and $C_{0}^{*}$ are the unique optimizers of 
\[
\left\{ \frac{b_{1}}{2}\left[m_{1}^{(2)}(c)C_{1}^{2}-m_{0}^{(2)}(c)C_{0}^{2}\right]\right\} ^{2}+\frac{\nu}{f(c)}\left\{ \frac{\sigma_{1}^{2}(c)}{C_{1}}+\frac{\sigma_{0}^{2}(c)}{C_{0}}\right\} ,
\]
with respect to $C_{1}$ and $C_{0}$. This implies that $\hat{C}_{1}$
and $\hat{C}_{0}$ also converge to the same respective limit $C_{1}^{*}$
and $C_{0}^{*}$ because the inequality will be violated otherwise.

Next we consider the case with $m_{1}^{(2)}(c)m_{0}^{(2)}(c)>0$.
In this case, with probability approaching 1, $\hat{m}_{1}^{(2)}(c)\hat{m}_{0}^{(2)}(c)>0$,
so that we assume this without loss of generality.

When these conditions hold, define $h_{0}=\hat{\lambda}_{2}h_{1}$
where $\hat{\lambda}_{2}=\{\hat{m}_{1}^{(2)}(c)/\hat{m}_{0}^{(2)}(c)\}^{1/2}$.
This sets the first-order bias term of $MMSE_{n}^{p}(h)$ equal to
0. Define the plug-in version of $AMSE_{2n}(h)$ by 
\[
AMSE_{2n}^{p}(h)=\Bigl\{\hat{b}_{2,1}(c)h_{1}^{3}-\hat{b}_{2,0}(c)h_{0}^{3}\Bigr\}^{2}+\frac{v}{n\hat{f}(c)}\left\{ \frac{\hat{\sigma}_{1}^{2}(c)}{h_{1}}+\frac{\hat{\sigma}_{0}^{2}(c)}{h_{0}}\right\} 
\]
Choosing $h_{1}$ to minimize $AMSE_{2n}^{p}(h)$, we define $\tilde{h}_{1}=\tilde{C}_{1}n^{-1/7}$
and $\tilde{h}_{0}=\tilde{C}_{0}n^{-1/7}$ where 
\begin{align}
\hat{\theta}_{2} & =\left\{ \frac{v\left[\hat{\sigma}_{1}^{2}(c)+\hat{\sigma}_{0}^{2}(c)/\hat{\lambda}_{2}\right]}{6\hat{f}(c)\left[\hat{b}_{2,1}(c)-{\hat{\lambda}_{2}}^{3}\hat{b}_{2,0}(c)\right]^{2}}\right\} ^{1/7}\quad\mbox{and}\quad\tilde{C}_{0}=\tilde{C}_{1}\hat{\lambda}_{2}.
\end{align}
Then $MMSE_{n}^{p}(\tilde{h})$ can be written as 
\[
MMSE_{n}^{p}(\tilde{h})=n^{-6/7}\left\{ \hat{b}_{2,1}(c)\tilde{C}_{1}^{3}-\hat{b}_{2,0}(c)\tilde{C}_{0}^{3}\right\} ^{2}+n^{-6/7}\frac{\nu}{\hat{f}(c)}\left\{ \frac{\hat{\sigma}_{1}^{2}(c)}{\tilde{C}_{1}}+\frac{\hat{\sigma}_{0}^{2}(c)}{\tilde{C}_{0}}\right\} .
\]

In order to match this rate of convergence, both $\hat{h}_{1}$ and
$\hat{h}_{0}$ need to converge at the rate slower than or equal to
$n^{-1/7}$ because the variance term needs to converge at the rate
$n^{-6/7}$ or faster. In order for the first-order bias term to match
this rate, 
\[
\hat{m}_{1}^{(2)}(c)\hat{h}_{1}^{2}-\hat{m}_{0}^{(2)}(c)\hat{h}_{0}^{2}\equiv B_{1n}=n^{-3/7}b_{1n},
\]
where $b_{1n}=O_{P}(1)$ so that under the assumption that $m_{0}^{(2)}(c)\neq0$,
with probability approaching 1, $\hat{m}_{0}^{(2)}(c)$ is bounded
away from 0 so that assuming this without loss of generality, we have
$\hat{h}_{0}^{2}=\hat{\lambda}_{2}^{2}\hat{h}_{1}^{2}-B_{1n}/\hat{m}_{0}^{(2)}(c)$.
Substituting this expression to the second term and the third term,
we have 
\begin{align*}
 & MMSE_{n}^{p}(\hat{h})=\left\{ \frac{b_{1}}{2}B_{1n}\right\} ^{2}+\left\{ \hat{b}_{2,1}(c)\hat{h}_{1}^{3}-\hat{b}_{2,0}(c)\{\hat{\lambda}_{2}^{2}\hat{h}_{1}^{2}-B_{1n}/\hat{m}_{0}^{(2)}(c)\}^{3/2}\right\} ^{2}\\
 & \hspace{65mm}+\frac{\nu}{n\hat{f}(c)}\left\{ \frac{\hat{\sigma}_{1}^{2}(c)}{\hat{h}_{1}}+\frac{\hat{\sigma}_{0}^{2}(c)}{\{\hat{\lambda}_{2}^{2}\hat{h}_{1}^{2}-B_{1n}/\hat{m}_{0}^{(2)}(c)\}^{1/2}}\right\} .
\end{align*}
Suppose $\hat{h}_{1}$ is of order slower than $n^{-1/7}$. Then because
$\hat{m}_{0}^{(2)}(c)^{3}\hat{b}_{2,1}(c)^{2}\neq\hat{m}_{1}^{(2)}(c)^{3}\hat{b}_{2,0}(c)^{2}$
and this holds even in the limit, the second-order bias term is of
order slower than $n^{-6/7}$. If $\hat{h}_{1}$ converges to 0 faster
than $n^{-1/7}$, then the variance term converges at the rate slower
than $n^{-6/7}$. Therefore we can write $\hat{h}_{1}=\hat{C}_{1}n^{-1/7}+o_{p}(n^{-1/7})$
for some $O_{P}(1)$ sequence $\hat{C}_{1}$ that is bounded away
from 0 and $\infty$ as $n\rightarrow\infty$ and as before $\hat{h}_{0}^{2}=\hat{\lambda}_{2}^{2}\hat{h}_{1}^{2}-B_{1n}/\hat{m}_{0}^{(2)}(c)$.
Using this expression, we can write 
\begin{align*}
 & MMSE_{n}^{p}(\hat{h})=n^{-6/7}\left\{ \frac{b_{1}}{2}b_{1n}\right\} ^{2}\\
 & \hspace{15mm}+n^{-6/7}\left\{ \left[\hat{b}_{2,1}(c)\hat{C}_{1}^{3}+o_{p}(1)-\hat{b}_{2,0}(c)\{\hat{\lambda}_{2}^{2}\hat{C}_{1}^{2}+o_{p}(1)-n^{-1/7}b_{1n}/\hat{m}_{0}^{(2)}(c)\}^{3/2}\right]\right\} ^{2}\\
 & \hspace{15mm}+n^{-6/7}\frac{\nu}{\hat{f}(c)}\left\{ \frac{\hat{\sigma}_{1}^{2}(c)}{\hat{C}_{1}+o_{p}(1)}+\frac{\hat{\sigma}_{0}^{2}(c)}{\{\hat{\lambda}_{2}^{2}\hat{C}_{1}^{2}+o_{p}(1)-n^{-1/7}b_{1n}/\hat{m}_{0}^{(2)}(c)\}^{1/2}}\right\} .
\end{align*}
Thus $b_{1n}$ converges in probability to 0. Otherwise the first-order
bias term remains and that contradicts the definition of $\hat{h}_{1}$.

Since $\hat{h}$ is the optimizer, $MMSE_{n}^{p}(\hat{h})/MMSE_{n}^{p}(\tilde{h})\leq1$.
Thus 
\[
\frac{o_{p}(1)+\left\{ \left[\hat{b}_{2,1}(c)\hat{C}_{1}^{3}-\hat{b}_{2,0}(c)\{\hat{\lambda}_{2}^{2}\hat{C}_{1}^{2}+o_{p}(1)\}^{3/2}\right]\right\} ^{2}+\frac{\nu}{\hat{f}(c)}\left\{ \frac{\hat{\sigma}_{1}^{2}(c)}{\hat{C}_{1}+o_{p}(1)}+\frac{\hat{\sigma}_{0}^{2}(c)}{\{\hat{\lambda}_{2}^{2}\hat{C}_{1}^{2}+o_{p}(1)\}^{1/2}}\right\} }{\left\{ \hat{b}_{2,1}(c)\tilde{C}_{1}^{3}-\hat{b}_{2,0}(c)\tilde{C}_{0}^{3}\right\} ^{2}+\frac{\nu}{\hat{f}(c)}\left\{ \frac{\hat{\sigma}_{1}^{2}(c)}{\tilde{C}_{1}}+\frac{\hat{\sigma}_{0}^{2}(c)}{\tilde{C}_{0}}\right\} }\leq1.
\]
If $\hat{C}_{1}-\tilde{C_{1}}$does not converge to 0 in probability,
then the ratio is not less than 1 at some point. hence $\hat{C}_{1}-\tilde{C_{1}}=o_{p}(1)$.
Therefore $\hat{h}_{0}/\tilde{h}_{0}\plim1$ as well.

The results shown above also imply that $MMSE_{n}^{p}(\hat{h})/MSE_{n}(h^{*})\plim1$
in both cases.\qed

\newpage													       
\section{Appendix}
\subsection{Introduction}
We present a detailed procedure to obtain RMSE* provided in Tables 1 and
2, a detailed algorithm to implement the proposed method, and a proof of
Lemma 1, in this supplemental material.

\subsection{A Procedure to Obtain RMSE*}
We describe how RMSE* is computed for the LLR estimators based on the
MMSE bandwidths, the IND bandwidths, and the IK bandwidth.
We also show how $\theta_{IK}$ in page 12 of the main text is obtained.

Once the sample size, the form of a kernel function, the functional
forms of $m_{1}(c)$, $m_{0}(c)$, $f(c)$, $\sigma_{1}^{2}(c)$,
and $\sigma_{0}^{2}(c)$ are given, the AMSE can be computed using the
formula of the AMSE in (2) for each of the bandwidths.

The MMSE bandwidths can be obtained by minimizing $MMSE_{n}(h)$ (not
$MMSE_{n}^{p}(h)$) provided in page 16 of the main text.
The IND bandwidths can be obtained based on the formulae provided in the footnote of page 12.

IK bandwidth can be obtained analogously except the regularization terms, $r_{+}+r_{-}$.
Note that 
\[
r_{+} = \frac{2160 \sigma_{1}^{2}(c)}{N_{2,+} h_{2,+}^{4}} \quad\mbox{and}\quad r_{-} = \frac{2160 \sigma_{0}^{2}(c)}{N_{2,-} h_{2,-}^{4}} 
\]
where
\[
h_{2,+} = 3.56 \left(\frac{\sigma_{1}^{2}(c)}{f(c) [m_{1}^{(3)}(c)]^{2}}\right)^{1/7} N_{+}^{-1/7}, \quad\mbox{and}\quad h_{2,-} = 3.56 \left(\frac{\sigma_{0}^{2}(c)}{f(c) [m_{0}^{(3)}(c)]^{2}}\right)^{1/7} N_{-}^{-1/7}.
\]
Hence the computation of the regularization term requires $N_{+}$,
 $N_{-}$, $N_{2,+}$, and $N_{2,-}$.
Since $N_{+}$ and $N_{-}$ are the number of observations to the left and right of the threshold, respectively (see p.942 of IK), their population analogues are computed by
\[
N_{+} = n\cdot \int_{-\infty}^{c}f(x)dx \quad\mbox{and}\quad N_{-} = n\cdot \int_{c}^{\infty} f(x) dx.
\]
Similarly, since $N_{2,+}$ and $N_{2,-}$ are the numbers of observations with $c\le X_{i} \le c +h_{2,+}$ and $c-h_{2,-} \le X_{i} <c$, respectively, their population analogues are computed by
\[
N_{2,+} = n\cdot \int_{c}^{c+h_{2,+}} f(x)dx \quad\mbox{and}\quad N_{2,-} = n\cdot \int_{c-h_{2,-}}^{c} f(x)dx.
\]
The same procedure is used to obtain $\theta_{IK}$ in page 12 in the main text.

\subsection{Implementation}\label{sec:Implementation}
To obtain the proposed bandwidths, we need pilot estimates of the
density, its first derivative, the second and third derivatives of the
conditional expectation functions, and the conditional variances at the cut-off point. 
We obtain these pilot estimates in a number of steps.\\

{\bf \large \noindent Step 1: Obtain pilot estimates for the density $f(c)$ and its first derivative $f^{(1)}(c)$}\\
We calculate the density of the assignment variable at the cut-off point, $f(c)$, which is estimated using the kernel density estimator with an Epanechnikov kernel.\footnote{IK estimated the density in a simpler manner (see Section 4.2 of IK). We used the kernel density estimator to be consistent with the estimation method used for the first derivative. Our unreported simulation experiments produced similar results for both methods.}
A pilot bandwidth for kernel density estimation is chosen using the normal scale rule with Epanechnikov kernel, given by $2.34 \hat \sigma n^{-1/5}$, where $\hat \sigma$ is the square root of the sample variance of $X_{i}$ (see \citealp{si86b} and \citealp{wj94} for the normal scale rules).
The first derivative of the density is estimated using the method proposed by \citet{jo94}. 
The kernel first derivative density estimator is given  by $\sum_{i=1}^{n} L((c-X_{i})/h)/(nh^{2})$, where $L$ is the kernel function proposed by \citet{jo94}, $L(u) =  -15 u (1-u^{2}) 1_{\{|u|<1\}}/4$. 
Again, a pilot bandwidth is obtained using the normal scale rule, given by $\hat \sigma \cdot (112\sqrt{\pi}/n)^{1/7}$.\\

{\bf \large \noindent  Step 2: Obtain pilot bandwidths for estimating the second and third derivatives $m_{j}^{(2)}(c)$ and $m_{j}^{(3)}(c)$ for $j=0, 1$}\\
We next estimate the second and third derivatives of the conditional mean functions using the third-order LPR.

We obtain pilot bandwidths for the LPR based on the estimated fourth derivatives of $m_{1}^{(4)}(c)=\lim_{x\to c+}m_{1}^{(4)}(x)$ and $m_{0}^{(4)}(c)=\lim_{x\to c-}m_{0}^{(4)}(x)$.
Following \citet{fg96}, \citet{ik12}, and \citet{cct14}, we use
estimates of $m_{1}^{(4)}(c)$ that are not necessarily consistent by fitting global polynomial regressions.
First, using observations for which $X_{i}\geq c$, we regress $Y_{i}$ on $1$, $(X_{i}-c)$, $(X_{i}-c)^{2}$, $(X_{i}-c)^{3}$ and $(X_{i}-c)^{4}$ to obtain the OLS coefficients $\hat\gamma_{1}$ and the variance estimate $\hat s_{1}^{2}$.
Using the data with $X_{i} < c$, we repeat the same procedure to obtain $\hat\gamma_{0}$ and $\hat s_{0}^{2}$.
The pilot estimates for fourth derivatives are $\hat m_{1}^{(4)}(c) = 24 \cdot \hat \gamma_{1}(5)$ and  $\hat m_{0}^{(4)}(c) = 24 \cdot \hat \gamma_{0}(5)$, where $\hat \gamma_{1}(5)$ and $\hat \gamma_{0}(5)$ are the fifth elements of $\hat \gamma_{1}$ and $\hat \gamma_{0}$, respectively.
The plug-in bandwidths for the third-order LPR used to estimate the second and third derivatives are calculated by
\[
h_{\nu,j} = C_{\nu,3}(K) \left(\frac{\hat s_{j}^{2}}{\hat f(c) \cdot \hat m_{j}^{(4)}(c)^{2} \cdot n_{j}}\right)^{1/9},
\]
where $j=0, 1$ (see \citealp[Section 3.2.3]{fg96} for information on
plug-in bandwidths and the definition of $C_{\nu,3}$).\footnote{The bandwidth we use
for estimating the third derivatives are not rate optimal when the
underlying function has higher order derivative.  However, we use this
bandwidth to avoid estimating higher order derivatives.}
We use $\nu=2$ and $\nu=3$ for estimating the second and third derivatives, respectively.\\

{\bf \large \noindent  Step 3: Estimation of the second and third derivatives $m_{j}^{(2)}(c)$ and $m_{j}^{(3)}(c)$ as well as the conditional variances $\hat\sigma_{j}^{2}(c)$ for $j=0, 1$}\\
We estimate the second and third derivatives at the cut-off point using the third-order LPR with the pilot bandwidths obtained in Step 2.
Following IK, we use the uniform kernel, which yields $C_{2,3} = 5.2088$ and $C_{3,3} = 4.8227$.
To estimate $\hat m_{1}^{(2)}(c)$, we construct a vector ${Y_{a}} = (Y_{1}, \ldots, Y_{n_{a}})'$ and an $n_{a}\times 4$ matrix, ${X_{a}}$, whose $i$th row is given by $(1, (X_{i}-c), (X_{i}-c)^{2}, (X_{i}-c)^{3})$ for observations with $c\le X_{i} \le c+h_{2,1}$, where $n_{a}$ is the number of observations with $c\le X_{i} \le c+h_{2,1}$.
The estimated second derivative is given by $\hat m_{1}^{(2)}(c) = 2 \cdot \hat\beta_{2,1}(3)$, where $\hat\beta_{2,1}(3)$ is the third element of $\hat\beta_{2,1}$ and $\hat\beta_{2,1} = ({X_{a}}'{X_{a}})^{-1}{X_{a}} {Y_{a}}$.
We estimate $\hat m_{0}^{(2)}(c)$ in the same manner.
Replacing $h_{2,1}$ with $h_{3,1}$ leads to an estimated third derivative of $\hat m_{1}^{(3)}(c) = 6 \cdot \hat\beta_{3,1}(4)$, where $\hat\beta_{3,1}(4)$ is the fourth element of $\hat\beta_{3,1}$, $\hat\beta_{3,1} = ({X_{b}}'{X_{b}})^{-1}{X_{b}} {Y_{b}}$,  ${Y_{b}} = (Y_{1}, \ldots, Y_{n_{b}})'$, ${X_{b}}$ is an $n_{b}\times 4$ matrix whose $i$th row is given by $(1, (X_{i}-c), (X_{i}-c)^{2}, (X_{i}-c)^{3})$ for observations with $c\le X_{i} \le c+h_{3,1}$, and $n_{b}$ is the number of observations with $c\le X_{i} \le c+h_{3,1}$.
The conditional variance at the cut-off point $\sigma_{1}^{2}(c)$ is
calculated as $\hat\sigma_{1}^{2}(c) = \sum_{i=1}^{n_{1}} (Y_{i} - \hat
Y_{i})^{2}/(n_1-4)$, where $\hat Y_{i}$ denotes the fitted values from
the regression used to estimate the second derivative.\footnote{Clearly,
the estimator is not a consistent estimator of the conditional variance,
but we do not need to estimate it consistently.  One can use a
non-parametric method to consistently estimate it, but it produces
almost identical simulation results.} $\hat \beta_{2,0}$, $\hat\beta_{3,0}$ and $\hat\sigma_{0}^{2}(c)$ can be obtained analogously.\\

{\bf \large \noindent  Step 4: Numerical Optimization}\\
The final step is to plug the pilot estimates into the $MMSE^{p}$ given by equation (8) in the main text and to use numerical minimization over the compact region to obtain $\hat h_{1}$ and $\hat h_{0}$.
Unlike $AMSE_{1n}(h)$ and $AMSE_{2n}(h)$ subject to the restriction given in Definition 1, the MMSE is not necessarily strictly convex, particularly when the sign of the product is positive.
In minimizing the objective function, it is important to try
optimization with several initial values,
in order to avoid finding only a local minimum.

\subsection{Proof of Lemma 1}
The LLR estimator can be expressed as $\left(\hat{\alpha}_{h_{1}}(c), \hat{\beta}_{h_{1}}(c) \right)' = \left(X(c)'W_{1}(c)X(c)\right)^{-1}X(c)'W_{1}(c)Y$,
where $X(c)$ is an $n\times 2$ matrix whose $i$th row is given by $(1, X_{i} -c)$, $Y=(Y_1,\ldots,Y_n)'$, $W_{1}(c) = {\rm diag}(K_{h_{1}}(X_{i}-c))$ and $K_{h_{1}}(\cdot) = K(\cdot/h_{1})\mathbb{I}\{\cdot\geq 0\}/h_{1}$.
The LLR estimator of $m_{1}(c)$ can also be written as $\hat\alpha_{h_{1}}(c) = e_1'\left(X(c)'W_{1}(c)X(c)\right)^{-1}X(c)'W_{1}(c)Y$,
where $e_1$ is a $2\times 1$ vector having one in the first entry and zero in the other entry.
Similarly, the LLR estimator for $m_{0}(c)$, denoted by $\hat \alpha_{h_{0}}(c)$, can be obtained by replacing $W_{1}(c)$ with $W_{0}(c)$, where $W_{0}(c) = {\rm diag}(K_{h_{0}}(X_{i}-c))$ and $K_{h_{0}}(\cdot) = K(\cdot/h_{0})\mathbb{I}\{\cdot < 0\}/h_{0}$.

A contribution to the MSE from a variance component is standard.
See \citet{fg96} for the details.
Here we consider the contribution made by the bias component.
We present the proof only for $\hat \alpha_{h_{1}}(c)$.
The proof for $\hat\alpha_{h_{0}}$ is parallel  and hence is omitted.
Denote $\hat \gamma_{1} =  \left(\hat\alpha_{h_{1}}(c),\; \hat\beta_{h_{1}}(c) \right)'$.
The conditional bias is given by
\[
\mbox{Bias}(\hat\gamma_{1}|X) = (X(c)'W_{1}(c)X(c))^{-1}X(c)W_{1}(c)(m_{1}-X(c)\gamma_{1}),
\]
where $m_{1} = (m_{1}(X_{1}),\ldots,m_{1}(X_{n}))'$ and $\gamma_{1} = (m_{1}(c), m_{1}^{(1)}(c))'$.
Define, for $j=0,1$ and an integer $k$,
\begin{align}
S_{n,k,j} &=\left[\begin{array}{ll} s_{n,k,j} & s_{n,k+1,j} \\ s_{n,k+1,j} & s_{n,k+2,j} \end{array}\right], \quad c_{n,k,j} = \left[\begin{array}{l} s_{n,k,j} \\ s_{n,k+1,j} \end{array} \right],\quad s_{n,k,j} = \sum_{i=1}^n K_{h_{j}}(X_{i}-c)(X_{i}-c)^{k},\nonumber\\
S_{k,1} &= \left[\begin{array}{cc} \mu_{k,0} & \mu_{k+1,0}\\ \mu_{k+1,0} & \mu_{k+2,0} \end{array} \right], \quad \mbox{and}\quad c_{k,1} = \left[ \begin{array}{c}
\mu_{k,0} \\ \mu_{k+1,0}
\end{array} \right]. \label{eq:bstdef}
\end{align}
Note that $S_{n,0,1}=X(c)'W_{1}(c)X(c)$.
The argument made by \citet{fghh96} can be generalized to yield
\begin{equation}\label{eq:snjb}
s_{n,k,1} = nh^{k}\left\{f(c)\mu_{k,0} + h f^{(1)}(c) \mu_{k+1,0} + o_{p}\left(h\right) \right\}.
\end{equation}
Then, it follows that
\[
S_{n,0,1} = n H \left\{ f(c) S_{0,1} + h f^{(1)}(c) S_{1,1} + o_{p}\left(h\right)\right\} H,
\]
where $H = \mbox{diag}(1,h)$.
By using the fact that $(A + hB )^{-1} = A^{-1} - h A^{-1} B A^{-1} + o\left(h\right)$, we obtain
\begin{equation}\label{eq:sinvb}
S_{n,0,1}^{-1} = n^{-1}H^{-1}\left\{ \frac1{f(c)} A_{0,1} - \frac{h f^{(1)}(c)}{f(c)^{2}} A_{1,1}  + o_{p}\left(h\right)\right\} H^{-1},
\end{equation}
where
\begin{align*}
A_{0,1} &= \left[\begin{array}{rr} \mu_{2,0} & -\mu_{1,0} \\ -\mu_{1,0} & \mu_{0,0}^{-1} \end{array}\right],\\
 A_{1,1} &= \frac1{\mu_{0,0}\mu_{2,0} - \mu_{1,0}^{2}}\left[\begin{array}{rr} -\mu_{1,0}(\mu_{2,0}^{2}-\mu_{1,0}\mu_{3,0}) & \mu_{2,0}(\mu_{2,0}^{2}-\mu_{1,0}\mu_{3,0}) \\ \mu_{2,0}(\mu_{2,0}^{2}-\mu_{1,0}\mu_{3,0}) & \mu_{1,0}^{3}-2\mu_{0,0}\mu_{1,0}\mu_{2,0} +\mu_{0,0}^{2}\mu_{3,0}  \end{array}\right].
\end{align*}

Next, we consider $X(c)W_{1}(c)\{m_{1}-X(c)\gamma_{1}\}$.
A Taylor expansion of $m_{1}(\cdot)$ yields
\begin{equation}\label{eq:beta2b}
X(c)W_{1}(c)\{m_{1}-X(c)\gamma_{1}\} = \frac{m_{1}^{(2)}(c)}{2} c_{n,2,1} + \frac{m_{1}^{(3)}(c)}{3!} c_{n,3,1} + o_{p}\left(nh^{3}\right).
\end{equation}
The definition of $c_{n,k,j}$ in (\ref{eq:bstdef}), in conjunction with (\ref{eq:snjb}), yields
\begin{equation}\label{eq:cnjb}
c_{n,k,1} = nh^{k}H\left\{ f(c) c_{k,1} + h f^{(1)}(c) c_{k+1,1} + o_{p}\left(h\right)\right\}.
\end{equation}
Combining this with (\ref{eq:sinvb}) and (\ref{eq:beta2b}) and extracting the first element gives
\[
\mbox{Bias}(\hat \alpha_{h_{1}}(c)|X) = \frac{ b_{1} m_{1}^{(2)}(c)}{2} h_{1}^{2} + b_{2,1}(c) h_{1}^{3} + o_{p} \left(h_{1}^{3} \right).
\]
This expression gives the required result.\qed

\bibliography{yarai}
\bibliographystyle{econometrica}
\end{document}